\documentclass[pre,10pt,showpacs]{revtex4}

\usepackage{amsmath}    
\usepackage{graphicx}   
\usepackage{verbatim}   
\usepackage{color}      
\usepackage{subfigure}  
\usepackage{hyperref}   
\usepackage{pstricks}
\usepackage{epstopdf}

\newcommand{\rf}[1]{Eq.~(\ref{#1})}

\def\etal{\textit{et al.~}}

\def\e{\textrm{e}}

\newcommand{\tr}[1]{\textrm{Tr}\left\{{#1}\right\}}

\def\ompl{\omega_\mathrm{pl}}

\usepackage{ifthen}
    \newboolean{showTODO}%
    \newboolean{showdetails}%
    \setboolean{showTODO}{false}%
    \setboolean{showdetails}{false}%

  \ifthenelse{\boolean{showdetails}}
             {
\newenvironment{details}[0]{%

\vspace*{0.02\textwidth}
\indent
\begin{minipage}[t]{0.80\textwidth}
\scriptsize
}{%
\end{minipage}%
\vspace*{0.02\textwidth}

} } {%
             }%

\begin{document}

\title{
Warm dense matter conductivity including electron-electron collisions
}

\author{H. Reinholz}
\affiliation{Universit\"at Rostock, Institut f\"ur Physik, 18051 Rostock, Germany}
\affiliation{University of Western Australia School of Physics, WA 6009 Crawley, Australia}

\author{G. R\"opke}
\affiliation{Universit\"at Rostock, Institut f\"ur Physik, 18051 Rostock, Germany}

\author{S. Rosmej}
\affiliation{Universit\"at Rostock, Institut f\"ur Physik, 18051 Rostock, Germany}

\author{R. Redmer}
\affiliation{Universit\"at Rostock, Institut f\"ur Physik, 18051 Rostock, Germany}

\email{heidi.reinholz@uni-rostock.de}

\date{\today}

\begin{abstract}
We present an approach that can resolve the controversy with respect to the role of electron-electron collisions in calculating the dynamic conductivity of dense plasmas. 
In particular, the dc conductivity is analyzed in the low-density, non-degenerate limit where the Spitzer theory is valid and 
electron-electron collisions lead to the well-known reduction in comparison to the result considering only electron-ion collisions (Lorentz model). 
With increasing degeneracy, the contribution of electron-electron collisions to the dc conductivity is decreasing 
and can be neglected for the liquid metal domain where the Ziman theory is applicable.
We give expressions for the effect of electron-electron collisions in calculating the conductivity in the warm dense matter region, i.e.\ for strongly coupled Coulomb systems at arbitrary degeneracy.
\end{abstract}

\maketitle

\section{Introduction}

Physical properties of warm dense matter (WDM) have become an emerging field of research.
New techniques such as intense ultra-short pulse laser irradiation or shock wave compression allow 
to produce states of matter with high energy density in the laboratory that are of relevance for astrophysical
processes. In the density-temperature plane of 
Coulomb systems, the region of degenerate, strongly coupled plasmas is now accessible.

The calculation of properties of WDM is a challenging task. Transport properties, in particular the dc conductivity,
are well investigated for a fully ionized plasma in the classical, low-density limit as given 
by Spitzer and H\"arm~\cite{Spitzer53} within kinetic theory (KT), see also~\cite{Brantov08} and references given there in.
The evolution of the electron velocity distribution function is described by a Fokker-Planck kinetic equation.
The linearized kinetic equations are solved with a Landau collision integral, 
that includes both the electron-ion ($e-i$) and electron-electron ($e-e$) collisions.  

Alternatively, the conductivity of strongly degenerate electron systems such as liquid metals has been
obtained by Ziman and Faber~\cite{Ziman} using the relaxation time approach.
The treatment of $e-i$ interaction has been improved by 
Dharma-wardana~\cite{dharma06} and others~\cite{Appel62,RH82,Chatto81} 
who used expressions for the pseudo-potentials and ionic structure factors that are appropriate for the particular 
ions under consideration. Lee and More~\cite{LeeMore} extended this approach to the non-degenerate regime. 
Desjarlais~\cite{MPD-LM} later derived corrections to the Lee-More conductivity model due to partial ionization.
However, to recover the Spitzer result for the conductivity, $e-e$ collisions have to be taken into account. 
This is not consistently possible within the relaxation time approach~\cite{Appel61}, but has been done by 
Stygar~\cite{stygar02} and Fortov \etal~\cite{Fortov03} using interpolation procedures, see also Adams \etal~\cite{Adams07}.
In this work, we present a general approach using linear response theory (LRT) that allows also for a 
systematic treatment of $e-e$ collisions at arbitrary degeneracy.

The investigation of time-dependent fields is somehow difficult in KT, too. 
Often,  the collision term in the time dependent kinetic equation is replaced by  an energy dependent but static relaxation time ansatz, 
see Landau and Lifshits~\cite{landau10}, Dharma-wardana~\cite{dharma06}, or Kurilenko~\etal~\cite{Berk92,Kuri95}. 
According to Landau and Lifshits \cite{landau10} it should be emphasized that such an approach 
is  only applicable in the low-frequency limit.   
The high-frequency region, relevant for describing bremsstrahlung, can be treated in LRT, see~\cite{Reinholz12}. 
In the present work, we focus on the static conductivity for a response to an electric field that is constant 
in time and space (dc conductivity).

Recently, the Kubo-Greenwood formula~\cite{Kubo66,Greenwood} was considered as a promising approach to the dynamical conductivity in dense, strongly interacting systems at arbitrary degeneracy. Based on the rich experience in electronic structure calculations for solids, liquids and complex molecules using density functional theory (DFT) and the enormous progress in computing power, \textit{ab initio} simulation techniques have been developed that allow to treat a large number of constituents with individual atomic structure. 
Most successful so far has been a combination of DFT for the electron system and classical molecular dynamics (MD) simulations for the ions which we will refer to as the DFT-MD method in what follows; for details, see~\cite{Desjarlais02,Mazevet05,holst08,Holst11}. This method does not rely on effective pair potentials or two-particle cross sections as in standard KT which become questionable in dense, strongly coupled plasmas. The evaluation of the Kubo-Greenwood formula using optimal single electron states gives the full account of $e-i$ interaction and treats $e-e$ interactions based on the \textit{exchange-correlation (XC)} functional used in the DFT cycle. The inclusion of $e-e$ \textit{collisions} into the DFT-MD calculations of transport properties in WDM is a subject of lively debate, especially for the limiting case of non-degeneracy. 

Within this paper, we apply  a generalized approach to non-equilibrium processes 
according to Zubarev \etal~\cite{ZMR2,Roepke13}. Using this generalized linear response theory (gLRT) 
transport properties are related to equilibrium correlation functions such as current-current or 
force-force correlation functions. 
Different expressions for the conductivity are deduced which lead to identical results  should  they be calculated exactly, 
as was shown analytically by performing partial integration. However, they are differently suited for performing calculations
after perturbation expansions. 
In particular, expressions that are consistent with KT (Spitzer result for the dc plasma conductivity)
are compared with the Ziman-Faber theory, the Kubo-Greenwood formula, and the rigorous results for 
the Lorentz model. In the Lorentz model, non-interacting electrons are considered to move under the 
influence of the potential of the ions at given configuration (adiabatic limit). 

Transport theory for WDM benefits from different sources. On one hand, the conductivity of liquid metals 
and disordered solids  is well described in the weak scattering limit (Fermi's golden rule) by the 
Ziman formula if the conducting electrons are degenerate, see also the Ziman-Faber approach~\cite{Ziman} where 
alloys at finite temperatures are considered~\cite{ChrisRoep85}. Main ingredients are the element-specific 
electron-ion pseudo-potential and the (dynamical) ion structure factor that are adequately described 
using the Kubo-Greenwood formula where the $e-i$ 
and the $e-e$ interaction (via the XC functional) are considered in any order. 
Evaluating the correlation functions within DFT-MD~\cite{Desjarlais02,Mazevet05,holst08,Holst11} no 
perturbation expansion is performed. On the other hand, the conductivity of plasmas is described by KT 
so that in the low-density, non-degenerate limit the Coulomb interaction between $e-i$ as well as $e-e$ 
pairs leads to the Spitzer result. At higher densities, gLRT can be applied that considers 
correlation functions to be evaluated analytically using the method of thermodynamic Green
functions~\cite{Roep88,Berk93a,rerrw99,Roepke13,RRN95}. Non-perturbative solutions are   possible
by classical MD simulations using effective pair potentials, see~\cite{Igor05}, as long as the non-degenerate case is considered. 

Bridging between both, the transport theory of condensed matter and plasma kinetic theory, 
the contribution of $e-e$ collisions that is clear in KT remains unclear in the Ziman or 
Kubo-Greenwood approach~\cite{dharma06}. We address this problem within gLRT that incorporates 
the Kubo formula as well as the KT as particular special cases in Sec.~\ref{sec:2}, see~\cite{Reinholz12,Roepke13}. 
A simple expression is derived that accounts for the contribution of $e-e$ interactions 
provided that the contribution of the $e-i$ interaction is known. Accounting for $e-e$ collisions, we show that the dc conductivity 
of WDM is reduced in the non-degenerate region what becomes 
less relevant with increasing degeneracy (Sec.~\ref{sec:dccond}). A simple fit formula is given in Sec.~\ref{correction}. Exemplarily, we present exploratory calculations for aluminum 
in the WDM region in Sec.~\ref{sec:aluminium}. Further properties such as the optical conductivity and general thermoelectric 
transport coefficients will be considered in subsequent work.

\section{Linear response theory and equilibrium correlation functions}
\label{sec:2}

\subsection{Fluctuations in equilibrium and transport properties}
\label{fluctuations}
In the following we outline the conceptional ideas on which the generalized response equations are based. 
The definitions of the physical system and the quantities for its description are given for completeness of the presentation.

We consider a charge-neutral Coulomb system consisting of ions with (effective) charge $Ze$ and particle density $n_{\rm ion}$, 
and electrons of charge $-e$, mass $m$, and particle density $n_{\rm e}=Zn_{\rm ion}$. The Hamiltonian 
\begin{equation}
\label{Hamilton}
\hat H=\hat T+\hat V_{\rm ei}+\hat V_{\rm ee} 
\end{equation}
of the system contains the kinetic energy ($\hat T$) of the electrons and ions,   
the electron-ion ($\hat V_{\rm ei}$) pseudo-potential and the electron-electron ($\hat V_{\rm ee}$) Coulomb 
interaction. 

The interaction with an external, spatially uniform electric field ${\bf E}^{\text{ext}}(t)$ is given by 
\begin{equation} \label{Hfield} 
\hat H_F(t)=-e \sum_i{\bf \hat r}_i \cdot {\bf E}^{\text{ext}}(t)
\end{equation}  
with ${\bf \hat r}_i$ the position operator of the different electrons in the considered sample. 
We take the adiabatic limit  and consider the electron contribution to the current density operator
\begin{equation} \label{curr}
{\bf \hat j}= \frac{e}{m \Omega}\sum_i {\bf \hat p}_i= \frac{e}{m \Omega} {\bf \hat P}\,,
\end{equation}  
$\Omega$ denotes the volume of the sample, and ${\bf \hat P}$ the total momentum of the electron subsystem. 
Without loss of generality we consider periodic time dependence of the field with frequency $\omega$. 
In LRT,  the average value of the current has the same periodic time dependence, 
$\langle {\bf \hat  j} \rangle^t= {\rm Re} [{\bf j}(\omega) \exp(-i \omega t)]$.
Similarly, an inhomogeneous external field can be decomposed into Fourier components with wave vector $\bf k$.
In the spatially homogeneous (${\bf k} \to 0$) and isotropic case considered here, the dynamical electric conductivity is defined as 
${\bf j}(\omega)= 
\sigma(\omega){\bf E}(\omega)$, 
where ${\bf E}(\omega)$ 
is the screened internal electric field.

There is a fundamental theory for transport coefficients that relates those to equilibrium correlation functions
\cite{ZMR2,Roepke13,Reinholz12}. We outline our approach and its general results in  App.~\ref{app:gLRT}. A main ingredient is the possibility to extend the relevant statistical operator considering a set $\{ \hat B_l \}$ of relevant observables that characterizes the non-equilibrium state of the system. 
The fluctuations of the single-particle occupation numbers or the respective current densities
could be considered. If the averages of these observables are already correctly taken into account, they don't have to be calculated dynamically so that the corresponding non-equilibrium state is observed within a shorter time when considering the evolution from an intitial state. 
As shown in App.~\ref{app:gLRT}, \textit{generalized response equations} are derived to eliminate the Lagrange parameters $F_n$ 
according to self-consistency conditions. Assuming linearity with respect to the external field, 
a system of linear equations follows where the coefficients are equilibrium correlation functions,
\begin{equation} \label{acf}
\left< \hat  A; \hat B \right>_z=\int_0^{\infty}  \textrm{d}t ~\e^{izt}\left( \hat  A(t), \hat  B\right)
=\int_0^{\infty}  \textrm{d}t ~\e^{izt}\int_0^1 {\rm d} \lambda~ \tr{ \hat A(t-i \hbar \beta\lambda)  \hat B^\dagger   \hat \rho_0 },
\end{equation}  
where $\rho_0$ is the equilibrium statistical operator. The time dependence  $ \hat A(t)=\e^{i \hat H t/\hbar} \hat A\e^{-i \hat H t/\hbar}$ is given by the Heisenberg 
picture with respect to the system Hamiltonian $ \hat H $,  so that ${\dot{ \hat A}}=i[  \hat H,  \hat A ]/\hbar $. $\beta = (k_{B}T)^{-1}$ is the inverse temperature.

\subsection{Different choices of relevant observables and corresponding response functions}
\label{subDiff}

Solving the \textit{generalized response equations},   transport coefficients are related to equilibrium correlation functions  which is an expression of the  fluctuation-dissipation theorem (FDT).
In principle, the equilibrium correlation functions (\ref{acf}) can be calculated because we know the equilibrium statistical operator. Thus, the FDT seems to be very convincing and 
promising to evaluate transport coefficients in dense, strongly correlated systems like WDM.
However, the evaluation of the equilibrium correlation functions is a quantum statistical many-body problem that has to 
be treated by 
perturbation theory or numerical simulations. For an analytical approach, the interaction between the charged
constituents of the system ($e,i$) is considered as perturbation. Additionally, we will show that the choice of  relevant observables $\{ \hat B_l \}$ is crucial for an effective solution scheme. 
We discuss three different  sets of relevant observables $B_l$ to characterize the non-equilibrium state, which are taken
in addition to the conserved observables energy $\hat H$ and particle number $\hat N$ of the system, see \cite{Roepke13}.

i) {\it The empty set of relevant  observables $ \hat B_l $ is considered}. It is equivalent to  the grand canonical ensemble, see Eq.~(\ref{rhorel}). 
All non-equilibrium distributions are formed dynamically. As the result we obtain the {\it Kubo formula}~\cite{Kubo66} 
\begin{equation}
\label{Kubo}
\sigma^{\rm Kubo}(\omega)=\frac{e^2 \beta }{3 m^2\Omega}\langle {\bf \hat P};{\bf \hat P} \rangle_{\omega+i \eta}^{{\rm irred}} \,,
\end{equation}
 where $\lim_{\eta \to 0}$ has to be taken after the thermodynamic limit. 
The response function is given  by  the correlation function of the electrical current, see Eq.~(\ref{curr}). 
It coincides with the conductivity $\sigma(\omega)$ if only the irreducible part of the current-current 
correlation function is taken. Despite this
compact, comprehensive and intuitive expression, 
its evaluation contains a number of difficulties. In particular, it is not suited 
for perturbation expansions of the dc conductivity because it is diverging in zeroth order of the interaction. We come back to this
issue in Sec.~\ref{subs:Pert} and App.~\ref{subsdeltaeta}.

ii)  {\it The fluctuations $\delta  \hat n_{\bf p} =\hat n_p-\langle \hat n_p \rangle_{\rm eq}$ of the single-particle occupation number $ \hat n_{\bf p}$ are chosen as relevant 
observables $B_l$.}  In this way, we can derive expressions in parallel to KT where 
the non-equilibrium state is characterized by the single-particle distribution function $f({\bf p},t)$. 
The modification  of the equilibrium single-particle distribution function can be calculated  
straightforwardly according to 
\begin{equation} \label{20}
\tr{ \hat \rho_{\rm rel}(t) \,\delta  \hat n_{\bf p}}=\sum_{{\bf p}'} \left(\delta  \hat n_{\bf p}, \delta  \hat n_{{\bf p}'}\right)  
 F_{{\bf p}'}(t) = \delta f({\bf p},t)\,.
\end{equation} 
The Lagrange multipliers $F_{\bf p}(t)$ are determined 
from the  response equations~(\ref{LBE1c}). These response equations are generalized linear {\it Boltzmann equations}
that contain a drift  and collision term as expressions of equilibrium correlation functions.  A comprehensive discussion is found in Ref.~\cite{Reinholz12}. 

The non-equilibrium single-particle distribution function (\ref{20}) 
is known if we have information about all moments of the distribution function, 
i.e.\ the quantum averages of the observables 
\begin{equation}
\label{moments}
 {\bf \hat P}_l=\sum_p {\bf p}\, \left(\frac{\beta\, {\bf p}^2}{2m}\right)^{(l-1)/2}\, \hat n_p \,. 
\end{equation} 
For instance, ${\bf \hat P}_1= {\bf \hat P}$ is related to the electrical current, and $ {\bf \hat P}_3$ to the heat current. 
Taking a finite number $L$ of these functions (\ref{moments}) as the set of relevant observables $\{B_l\}$, see \cite{Redmer97,Reinholz05,Reinholz12,Roepke13}, the response function is approximated by a ratio of two determinants  
\begin{equation}
\label{sigma} 
\sigma^{(L)}(\omega) = - \frac{e^2 \beta}{ m^2 \Omega}
\begin{vmatrix} 0& N_{11}&\ldots & N_{1L}\\
              N_{11} & d_{11}&\ldots & d_{1L}\\
              \vdots& \vdots& \ddots& \vdots&\\ 
              N_{L1} & d_{L1}&\ldots & d_{LL}
\end{vmatrix}
/\begin{vmatrix} 
               d_{11}&\ldots & d_{1L}\\
               \vdots& \ddots& \vdots&\\ 
               d_{L1}&\ldots & d_{LL}
\end{vmatrix}.
\end{equation} 
The Kubo scalar products are given analytically for the electron gas as
\begin{equation}
\label{N}
N_{ll'}=\frac{1}{3}({\bf \hat P}_l,{\bf \hat P}_{l'})=\frac{Zn_{\rm ion}\, \Omega \,m}{\beta}\frac{\Gamma((l+l'+3)/2)}{\Gamma(5/2)}\frac{I_{(l+l'-1)/2}(\beta\mu_{e}^{id})}{I_{1/2}(\beta\mu_{e}^{id})},
\end{equation}
with the ideal part of the electron chemical potential $\mu_{e}^{id}$ and the 
Fermi integrals $I_{\nu}(y)=\frac{1}{\Gamma(\nu+1)}\int\limits_{0}^{\infty}\frac{x^{\nu}dx}{e^{x-y}+1}$.
\begin{equation} \label{dnm}
d_{ll'}(\omega) =\frac{1}{3} \{ \langle {\bf \dot {\hat P}}_l ;{\bf \dot {\hat P}}_{l'} \rangle_{\omega+i\eta}^{{\rm irred}} -i\omega ({\bf \hat P}_l,{\bf \hat P}_{l'}) \} 
\end{equation}
are correlation functions (\ref{acf}) of the system in thermodynamic equilibrium. 

With increasing number of moments, $L \to \infty$, the full solution of the KT 
would be reproduced. 
Further convergence issues, in particular for the static case, have been discussed in detail elsewhere, 
see~\cite{RRN95,Redmer97,RRT89,Karachtanov13,Roepke89}.
Note that the static conductivity is increasing if more moments are taken into account as a consequence 
of the Kohler variational principle, see Ref.~\cite{Reinholz12}. 

iii)  { \it The current density operator is taken as relevant observable $ \hat B_l$}.  
This relates directly to the thermodynamics of irreversible processes where the state of the system is  described by currents. 
Corresponding generalized forces  are identified as the response parameters  $ F_l$. 
In a first step, we consider the total momentum ${\bf \hat P}$ as relevant observable. The relevant distribution function
is a shifted Fermi or Boltzmann distribution.
Further details of the non-equilibrium distribution functions beyond the average of momentum (which is correctly reproduced) are formed dynamically. We obtain
\begin{equation}
\label{FF}
\sigma^{\rm Ziman}(\omega)=\frac{e^2\beta}{3 m^2\Omega}\frac{({\bf \hat P},{\bf \hat P})^2}{-i \omega ({\bf \hat P},{\bf \hat P})+
\langle { {\bf \dot {\hat P}} ;{\bf \dot {\hat P}} }\rangle_{\omega+i\eta}^{{\rm irred}}}
\end{equation}
which we denote by Ziman since its static limit ($\omega=0$) for $T=0$~K is the Ziman-Faber formula \cite{Ziman} for the  conductivity. The Ziman formula is also denoted as second fluctuation-dissipation theorem 
since the inverse transport coefficients are related to the force-force correlation function 
$\langle { {\bf \dot {\hat P}} ;{\bf \dot {\hat P}} }\rangle_{\omega+i\eta}$. 
Note that ${\bf \hat P}$ is the first moment of the single-particle distribution function (\ref{moments}). 
Therefore, the response function (\ref{FF}) is identical with Eq.~(\ref{sigma}) for $L=1$, the first moment approach in KT. 

Using the explicit expression for the Kubo scalar product (\ref{N}), we obtain from Eq.~(\ref{FF})
a generalized Drude expression for the conductivity~\cite{rerrw00}
\begin{equation}\label{sigmaom}
 \sigma^{\rm Ziman}(\omega)=\frac{\epsilon_0 \ompl^2}{-i\omega + \nu^{\rm Ziman}(\omega)} \,,
\end{equation}
with the plasma frequency $\ompl=\sqrt{e^2 Zn_{\rm ion}/(\epsilon_0 m)}$. The dynamical collision frequency 
\begin{equation} \label{nuZiman}
 \nu^{\rm Ziman}(\omega)= \frac{\beta}{3Z n_{\rm ion}\Omega m}\langle { {\bf \dot {\hat P}} ;{\bf \dot {\hat P}} }\rangle_{\omega+i\eta}^{\textrm{irred}}
\end{equation} 
is given in terms of the irreducible part  of the force-force correlation function. 
However, higher moments are needed in order to take into account $e-e$ 
collisions. As a special case, the two-moment approach with ${\bf \hat P}_1, {\bf \hat P}_3$ as relevant observables 
is discussed in Ref.~\cite{Reinholz12} and will be considered in the explicit calculations in Sec.~\ref{sec:dccond}.

\subsection{Perturbation theory for the dynamic conductivity and convergence}
\label{subs:Pert}

For the dynamic conductivity, we derived expressions  (\ref{Kubo}), (\ref{sigma}) and
(\ref{FF}) 
which can be proven to be identical by performing partial integration, see~\cite{Reinholz12,Roepke13}. 
They have, however, different properties when considering the time behavior of the respective equilibrium 
correlation functions and systematic perturbation expansions. This will be discussed in the following.

\textit{Version (i), the Kubo formula ~(\ref{Kubo}) and the current-current correlation function.} 
The momentum of the electrons is conserved in zeroth order of the 
interaction with the ions. Explicitly, evaluating the  correlation function (\ref{acf})  in lowest order, we have
\begin{equation}
 \sigma^{\rm Kubo}(\omega) =\lim_{\eta \to 0}\frac{e^2 \beta}{3 m^2 \Omega} \frac{({\bf \hat P},{\bf \hat P})}{-i \omega + \eta}
=\lim_{\eta \to 0}\frac{\epsilon_{0}\omega_{\rm pl}^2}{-i \omega + \eta}.
\end{equation}
 The dynamical conductivity is purely imaginary for finite frequencies. 
This result is well-known as the Lindhard RPA expression of the dielectric function $\epsilon(0,\omega) = 1 - \omega_{\rm pl}^2/\omega^2$. The dc conductivity ($\omega \to 0$) diverges.
Therefore, the Kubo formula~(\ref{Kubo}) is not appropriate to calculate the dc conductivity within perturbation theory. 
Applying a perturbation expansion, additional steps like partial summations or $\delta_\eta$ functions with finite width 
are required, see Appendix \ref{subsdeltaeta}. 
Note, that perturbation theory is suitable for finite frequencies. 

\textit{Version (ii), the kinetic theory and the single-particle occupation number correlation function.} 
The correlation functions $d_{ll'}$ (\ref{dnm}) in the expression for the conductivity (\ref{sigma}) can be evaluated  
by perturbation theory using thermodynamic Green's functions, 
see~\cite{Reinholz12} and Sec. \ref{sec:r}. From the definition of the generalized forces 
${\bf \dot {\hat P}}_l=i[\hat H,{\bf \hat P}_l]/\hbar$ with $\hat H$ containing kinetic and potential energy, see Eq.~(\ref{Hamilton}), it is evident that $d_{ll'}(\omega =0)$ is of second order in the interaction. 
The kinetic energy $\hat T$ commutes with ${\bf \hat P}_l$.  
The quantity ${\bf \dot {\hat P}}_l$ entering the correlation function  $d_{ll'}$ is decomposed in the 
contributions due to the $e-i$ and the $e-e$ interaction. The evaluation of the correlation functions 
\begin{equation}
\label{d}
\langle {\bf \dot {\hat P}}_l ;{\bf \dot {\hat P}}_{l'} \rangle_{\omega+i\eta}
=-\frac{1}{\hbar^2}\left\{ \langle [\hat V_{\rm ei},{\bf \hat P}_l];[\hat V_{\rm ei},{\bf \hat P}_{l'}] \rangle_{\omega+i\eta}
+\langle [\hat V_{\rm ee},{\bf \hat P}_l];[\hat V_{\rm ee},{\bf \hat P}_{l'}] \rangle_{\omega+i\eta}\right\}
\end{equation}
in Born approximation for the screened Coulomb  potential $\hat V_{\rm ei}$ is given  in Sec.~\ref{sec:r} for the static case; 
 for arbitrary $\omega$ see~\cite{Reinholz12}. It should be emphasized  that for the dc conductivity a 
perturbation expansion is possible starting with a non-diverging term in lowest order, in contrast to 
the Kubo formula~(\ref{Kubo}). Contributions due to $e-e$ collisions are represented by the second term 
in Eq.~(\ref{d}) for $l,l' > 1$ only since the lowest-order term vanishes, 
$[{\bf \hat P}_{1},\hat V_{\rm ee}]=0$.

\textit{Version (iii), the Ziman formula~(\ref{FF}) and the force-force correlation function.} 
Following the discussion of the correlation functions $d_{ll'}$ (\ref{d}) it is evident that 
the collision frequency $\nu^{\rm Ziman}(\omega)$, Eq.~(\ref{nuZiman}), behaves regular 
in the limit $\omega \to 0$ so that one can also perform this limit in expression (\ref{sigmaom}). 
However, since the $e-e$  interaction does not contribute to ${\bf \dot {\hat P}} \equiv {\bf \dot {\hat P}}_1$, 
it treats the conductivity on the level of the Lorentz model only. It does not give 
the correct result in the low-density limit as was discussed in \cite{Reinholz12}, 
see also the following section \ref{sec:dccond}, but is correct  in the limit of strong degeneracy.

The well known expression of the Ziman formula for $\omega=0$ was derived in Born approximation.  
It can be improved considering higher-order terms in the perturbation 
expansion~\cite{ChrisRoep85}. However, then also secular divergent terms (van Hove limit) arise 
that have to be treated by partial summations~\cite{AS74,HC75}. This is avoided if the single-particle 
distribution function is considered as relevant observable. The account of higher 
moments ${\bf \hat P}_l$ of the single-particle distribution function also improves the result for the 
Born approximation. For increasing numbers of moments, see~\cite{RRN95,RRT89}, the solution converges to the Spitzer formula if considering the low-density limit.
Going beyond ${\bf \hat P}_1$, the $e-e$ collisions contribute. 
Thus, to avoid singular expansions and partial summations, we can enlarge the number of relevant observables 
corresponding to the Kohler variational principle as given by version (ii), see~\cite{Reinholz12}.

\section{Dc conductivity and electron-electron collisions} \label{sec:dccond}

\subsection{Renormalization function}
\label{sec:r}

As was shown in the previous section, the best choice of relevant observables to take into account $e-e$ collisions are the fluctuations of the single-particle occupation numbers, leading to Eq.~(\ref{sigma}). The calculations can be performed in Born approximation without encountering any divergencies and naturally including all relevant scattering mechanisms.
Adopting the  Drude form (\ref{sigmaom}) obtained from the Ziman formula as the general expression for the conductivity, 
\begin{equation}\label{sigmaom1}
 \sigma(\omega)\equiv\frac{\epsilon_0 \ompl^2}{-i\omega + \nu(\omega)} \,,
\end{equation}
we  take this as definition of the dynamical collision frequency $\nu(\omega)$. 
To show the influence of $e-e$ collisions on the conductivity we  relate the \textit{full} dynamical collision frequency to the  
solution in the one-moment approach (\ref{nuZiman}) as a reference value by  introducing a complex renormalization function $r(\omega)$ in such a way that  
\begin{equation}\label{sigmaom2}
 \nu(\omega)\equiv r(\omega)  \nu^{\rm Ziman}(\omega)= 
 r(\omega) \frac{1}{({\bf \hat P},{\bf \hat P})} \langle { {\bf \dot {\hat P}} ;{\bf \dot {\hat P}} }\rangle_{\omega+i\eta}^{\rm irred} \,.
\end{equation}
If the solution is approximated within a finite number $L$ of moments according to Eq. (\ref{sigma}), a  
renormalization function  $r^{(L)}(\omega)$ is defined correspondingly so that, 
see Refs.~\cite{rerrw00,Reinholz05,rr00},
\begin{equation}\label{Drude2}
 \sigma^{(L)}( \omega)=\frac{\epsilon_0 \ompl^2}{-i\omega + r^{(L)}(\omega)\nu^{\rm Ziman}(\omega)}\,.
\end{equation} 

Let us consider the simplest non-trivial approximation, the two-moment approach  with ${\bf \hat P}_1, {\bf \hat P}_3$, 
i.e.~particle current and energy current as relevant observables. Then, from Eq.~(\ref{sigma}),
the renormalization factor can be given explicitly in the static (dc) case as (for the dynamic case, see \cite{Reinholz12})
\begin{equation}\label{3sigma}
r^{(2)}(0) =  
\frac{ d_{33} d_{11} - d_{13} d_{31}} {  d_{11}   
\left[ d_{33}  +\frac{N_{13}^2}{N_{11}^2}d_{11}
-\frac{N_{31}}{N11}d_{13}-\frac{N_{13}}{N11}d_{31}  \right]}\,.
\end{equation}

The correlation functions 
$d_{ll'}=\frac{1}{3}\left< { \bf \dot  {\hat P}}_{l};{ \bf \dot  {\hat P}}_{l'} \right>_{i\epsilon}=d_{ll'}^{\rm ei}+d_{ll'}^{\rm ee}$ 
have to be evaluated. 
The non-degenerate limit for a plasma with singly charged ions has already been 
discussed in~\cite{Reinholz12}. Here we will calculate the renormalization function for arbitrary degeneracy  and  effective ion charge $Z$.
In screened Born approximation, we have (summation over $k,p$ includes spin and respective wave vector summation)
\begin{eqnarray}
\label{deiLB}
d_{ll'}^{\rm ei} &=& \pi\hbar Z^2\sum_{{k},{p},{q}}\int\limits_{-\infty}^{\infty}d \hbar \omega
\left|\frac{V(q)}{\epsilon^{\rm RPA}(q,\omega)}\right|^{2}f_{k}^{e}(1-f_{\left|\bf{k}+\bf{q}\right|}^{e}) 
f_{p}^{i}(1-f_{\left|\bf{p}-\bf{q}\right|}^{i}) \\ \nonumber
 &&\times \delta(\hbar \omega -E_{\left|\bf{k}+\bf{q}\right|}^{e}+E_{k}^{e})\delta(\hbar \omega -E_{p}^{i}+E_{\left|\bf{p}-\bf{q}\right|}^{i}) 
 K_{l}({\bf k},{\bf q})K_{l'}(\bf{k},\bf{q}) \,,
\\ && \nonumber \\
\label{deeLB}
d_{ll'}^{\rm ee} &=& \frac{\pi\hbar}{2} \sum_{{k},{p},{q}}\int\limits_{-\infty}^{\infty}d \hbar \omega
\left|\frac{V(q)}{\epsilon^{\rm RPA}(q,\omega)}\right|^{2}
f_{k}^{e}(1-f_{\left|\bf{k}+\bf{q}\right|}^{e}) f_{p}^{e}(1-f_{\left|\bf{p}-\bf{q}\right|}^{e})
 \\ && \times \delta(\hbar \omega -E_{\left|\bf{k}+\bf{q}\right|}^{e}+E_{k}^{e})
\delta(\hbar \omega -E_{p}^{e}+E_{\left|\bf{p}-\bf{q}\right|}^{e}) 
(K_{l}({\bf k},{\bf q})+K_{l}({\bf p},-{\bf q}))(K_{l'}({\bf k},{\bf q})+K_{l'}({\bf p},-{\bf q})) \,, \nonumber
\end{eqnarray}  
where $f_{k}^c=(e^{\beta(E_k^c-\mu_c^{id})}+1)^{-1}$, 
$E_{k}^{c}=\hbar^{2}k^{2}/(2m_{c})$ and 
$K_{l}({\bf k},{\bf q})=k_{z}(\beta E_{k}^{e})^{(l-1)/2}-(k_{z}+q_{z})(\beta E_{\left|\bf{k}+\bf{q}\right|}^{e})^{(l-1)/2}$ 
with index $c=i,e$ for ion and electron contributions, respectively.
Exchange terms in $d_{ll'}^{\rm ee}$ are small and not given here. The Coulomb interaction 
$V(q) = e^2/(\epsilon_0 \Omega q^2)$ is statically screened 
with $\epsilon^{\rm RPA}(q,0)=1+\kappa^2/q^2$, where
$\kappa^2= (2\Lambda_e^{-3} I_{-1/2}(\beta \mu_e^{id})+Z^{2} n_{\rm ion} ) e^2/(\epsilon_0 k_BT)$  
is related to the Debye screening length, the (ideal) electron chemical potential $\mu_e^{id}$ (see Eq.~(\ref{N})) 
and thermal wavelength $\Lambda_e=(2\pi\hbar^2/m k_BT)^{1/2}$. 
The explicit evaluation of the correlation functions 
$d_{ll'}^{\rm ei},d_{ll'}^{\rm ee}$  in Born approximation relevant for the two-moment approach $r^{(2)}(0)$,
Eq.~(\ref{3sigma}), is shown in App.~\ref{app0}. 
The limit of non-degenerate electrons is discussed in the following subsection. 

Whereas  the statically screened Coulomb potential  is a reasonable description for the $e-e$ interaction
leading to a convergent result for the correlation function, taking this approximation for the interaction of electrons with ions (effective charge $Z$)  is only applicable in the low-density limit.
For WDM at higher densities, the interaction at short distances is of relevance where the Coulomb potential 
has to be replaced by a pseudo-potential. Also, the ionic contribution to the screening should be taken into account via
the ion-ion structure factor. Both effects are taken into account in DFT-MD simulations, see Sec.~\ref{sec:KG}.
They would improve the result for $d_{ll'}^{\rm ei}$ in the high-density region.

By introducing the renormalization function $r^{L}(\omega)$ in Eq.~(\ref{Drude2}) we have improved the Ziman result for the conductivity to the full  solution of KT if an infinite set of moments is used, $L\to\infty$. Furthermore, evaluation of the correlation functions (\ref{deiLB}) and (\ref{deeLB}) allows considering the influence of $e-e$ collisions beyond the Lorentz model so that the correct Spitzer result is obtained in the low-density limit. For the following discussions it is helpful to introduce a correction 
factor, see also~\cite{Adams07},
\begin{equation}\label{eq:Ree}
 R_{\rm ee}(\omega)= \frac{\sigma_{\rm ei+ee}(\omega)}{\sigma_{\rm ei}(\omega)} \,,
\end{equation}
where ${\sigma}_{\rm ei+ee}(\omega)$ denotes the dynamical conductivity determined within gLRT including the $e-e$ interaction, whereas ${\sigma}_{\rm ei}(\omega)$ is that of the Lorentz model neglecting $e-e$ interactions.

\subsection{Non-degenerate plasma with singly-charged ions} \label{subsec:limits}

Here we discuss results for the fully ionized hydrogen plasma ($Z=1$) in the low-density limit, 
see~\cite{Redmer97,Roep88,EssRoep98,RR89}. We introduce the plasma parameter 
$\Gamma = e^2 (4\pi n_{\rm e}/3)^{1/3}/(4\pi\epsilon_0 k_BT)$ 
and the electron degeneracy parameter 
\begin{equation}\label{theta}
\Theta = \frac{2m k_BT}{\hbar^2} (3\pi^2 n_{\rm e})^{-2/3}
\end{equation}
 as dimensionless parameters. 
Due to simple dependencies in the low-density limit ($\Gamma \ll 1, \Theta \gg 1$),  
the dc conductivity $\sigma(n_{\rm e},T)$ is traditionally also  related to a dimensionless function 
$\sigma^*(\Gamma,\Theta)$ according to 
\begin{equation}
 \sigma(n_{\rm e},T)=\frac{(k_BT)^{3/2} (4 \pi \epsilon_0 )^2}{m^{1/2} e^2}\;\sigma^*(\Gamma,\Theta) \,.
\end{equation}
In the low-density limit, this function can be expressed as
\begin{equation} \label{lowdenslimit}
 \sigma^*(\Gamma,\Theta)= \frac{\textrm{prefactor} \; a} {\textrm{Coulomb logarithm} \;L(\Gamma,\Theta)} \,.
\end{equation}
Explicit expressions of the Coulomb logarithm $L$ depend on the treatment of the collision term, in particular 
the  screening and whether strong collisions have been taken into account, see~\cite{Redmer97,Roep88,RR89}. Different approximations and approaches are summarized in Tab.~\ref{tabL}.

\begin{table}[h] 
\caption{Coulomb logarithm and prefactor in the low-density limit according to Eq.~(\ref{lowdenslimit}) 
for different approximations originally derived via the Fokker-Planck equation (FP), the relaxation time approximation (RTA), 
or linear response theory (LRT). Collisions are treated in Born approximation (weak, B) or T-matrix (strong, T).}  
\label{tabL}
\begin{center}
\begin{tabular}{lccccccc} 
\hline\hline
& notation & collisions & originally & \multicolumn{2}{c}{prefactor $a$} &  Coulomb \\ 
& & & derived by & $ei$ & $ei+ee$ & logarithm \\
\hline
Spitzer~\cite{Spitzer53} & $\sigma^{\rm KT}$ & strong & FP & 1.016 & 0.591 & $L_{\rm Sp}$  \\ 
Brooks-Herring~\cite{Brooks} & $\sigma^{\rm Lorentz}$ & weak & RTA & 1.016 & - & $L_{\rm BH}$  \\ 
Ziman~\cite{Ziman} & $\sigma^{\rm Ziman}$ & weak & RTA & 0.299 & - & $L_{\rm Zi}$  \\ 
Eq.~(\ref{sigma}), 1 moment  & $\sigma^{\rm (1),B}$ & weak & LRT & 0.299 & - &  $L_{\rm Zi}$  \\ 
Eq.~(\ref{sigma}), 2 moments & $\sigma^{\rm (2),B}$ & weak & LRT & 0.972 & 0.578 &  $L_{\rm Zi}$  \\ 
Eq.~(\ref{sigma}), 2 moments & $\sigma^{\rm (2),T}$ & strong & LRT & 0.972 & 0.578 & $L_{\rm Sp}$  \\ \hline\hline
\end{tabular}
\end{center}
\end{table}

KT for the fully ionized plasma in the high-temperature, low-density limit leads to the Spitzer result~\cite{Spitzer53} 
with the Spitzer Coulomb logarithm 
\begin{align}\label{eq:Spitzer1}
  L_{\rm Sp}(\Gamma)=\frac{1}{2} \ln \left(\frac{3}{2} \Gamma^{-3}\right) ,
\end{align}
valid for $\Gamma^{2} \Theta \gg 1$ only. Strong collisions as well as $e-e$ collisions are taken into account. 
Contrary, the relaxation time approximation allows the derivation of analytical expressions for $\Gamma^{2} \Theta \ll 1$ 
in the case of the Lorentz plasma valid for highly charged ions where collisions can be treated within Born approximation. 
What follows is the Brooks-Herring formula~\cite{Brooks} 
with the Brooks-Herring Coulomb logarithm 
\begin{align}\label{eq:Brooks1}
 L_{\rm BH}(\Gamma, \Theta)=-\frac{1}{2} \ln(\zeta)-\frac{1}{2} (\gamma+1)-\zeta \ln(\zeta)+\dots,
\end{align}
where 
$\zeta=(2/3\pi^2)^{1/3} \Gamma/\Theta$, and $\gamma=0.577216\dots$ is Euler's constant. 

For completeness, we give the Ziman formula~\cite{Ziman} that arises from evaluating the force-force correlation function $d_{ll'}^{\textrm{ei}}$ (\ref{deiLB}), calculating the Born approximation in the adiabatic and static case. For the Coulomb logarithm we have
\begin{align}\label{Ziman}
  L_{\rm Zi} = \frac{3\pi^{1/2}}{4} \Theta^{3/2} \int_0^\infty dq\,\, q^3 f_e(q/2) \left| \frac{V_{\rm ei}(q)}{\epsilon^{\rm RPA}_e(q,0)}\right|^2 \frac{\epsilon_0^2}{e^4} \, S_{ii}(q) \,,
\end{align}
containing the static ion-ion structure factor $S_{ii}(q)$. 
This Coulomb logarithm is applicable for any degeneracy and leads to the Brooks-Herring Coulomb logarithm in the low-density limit 
($\Theta \gg 1, S_{ii}(q)=1$).

Inspecting Tab.~\ref{tabL}, it is apparent that the known limiting cases discussed above can be reproduced within gLRT. 
While the one-moment approximation leads to the Ziman formula, we conclude that the two-moment approach is already a reasonable 
approximation to the prefactor $a$ given by Spitzer. It can be improved taking higher moments into account~\cite{RRN95,RRT89}. 

In any case, we find the low-density limit $L = -\frac{1}{2} \ln(n) +{\cal O}(n^0)$,
where the contributions ${\cal O}(n^0)$ depend on the plasma parameters and the approximation taken. 
Regardless of the treatment of the collision integral, the conductivity is lower if $e-e$ collisions 
are taken into account. 
For the fully ionized hydrogen plasma ($Z=1$) in the low-density limit, we find the correction factor (\ref{eq:Ree}) 
\begin{equation} \label{eq:ReeKT}
 R_{\rm ee}^{\rm KT}=\lim_{\Theta \gg 1} \frac{{\sigma}^{\rm KT}}{{\sigma}^{\rm Lorentz}} = \frac{0.591}{1.016}=0.582 \,,
\end{equation}
from the prefactors given in Tab.~\ref{tabL}.  
The prefactor $a^{\rm Lorentz}=2^{5/2} \pi^{-3/2} \approx 1.016$ results from solving the Fokker-Planck equation for  the Lorentz model. The same prefactor is found in the Brooks-Herring formula~(\ref{eq:Brooks1}) using the relaxation time ansatz. 
Thus, this result corresponds to the evaluation of the conductivity according to version (ii) by taking into account 
arbitrary numbers of moments. 

The Ziman formula~(\ref{Ziman}) with the prefactor $a^{\rm Ziman} = 3/[4 (2 \pi)^{1/2}]\approx 0.299$  can be applied 
to the strongly degenerate electron gas, but is no longer exact for higher temperatures. As discussed above, the 
force-force correlation function [version (iii) in Subsec. \ref{subs:Pert}]  in Born approximation cannot reproduce the details of the distribution function. 
The inclusion of $e-e$ scattering leads to the prefactor 0.591 in the Spitzer formula~(\ref{eq:Spitzer1}); 
this result is reproduced starting from Eq.~(\ref{sigma}). The convergence with increasing rank $L$ is shown, for instance, 
in~\cite{Redmer97,RRN95,RRT89}. We now compare the exact limit Eq.~(\ref{eq:ReeKT}) with the result using the prefactors 
in the two-moment approach.
\begin{equation} \label{eq:R(2)}
 R_{\rm ee}^{(2)}=\lim_{\Theta \gg 1} \frac{\sigma^{(2)}_{\rm ei+ee}}{\sigma^{(2)}_{\rm ei}} = \frac{0.578}{0.972}=0.594 \,.
\end{equation}

The two-moment approach with ${\bf P}_1,\, {\bf P}_3$ as relevant observables (i.e.\ 
particle current and energy current) allows for a variational approach to the 
single-particle distribution function working well for the low-density, 
non-degenerate limit. It will be extended to arbitrary degeneracy in Subsec.~\ref{correction}.

\subsection{The Kubo-Greenwood formula: DFT-MD calculations of correlation functions in WDM} \label{sec:KG}

Recent progress in numerical simulations of many-particle systems allows to calculate 
correlation functions in WDM, e.g.~in planetary interiors~\cite{French12}. 
In classical systems, MD simulations have been performed for sufficiently large 
systems using effective two-particle potentials in order to obtain correlation 
functions that can be compared with analytical results, see~\cite{Igor05,Baalrud13,Lorin15}. 
In WDM, it is inevitable to allow for quantum effects and strong correlations 
in the region where electrons are degenerate. This can be done, for example, within MD 
simulations based on finite-temperature DFT using Kohn-Sham (KS) single-electron states. 
To treat a disordered system of moving ions in adiabatic approximation, 
in addition to the general periodic boundary conditions for the macroscopic system, 
the ion positions are fixed in a finite supercell (volume $\Omega_c$) at each time step
so that the KS potential is periodic with respect to this supercell. We can introduce Bloch 
states $u_{ {\bf k} \nu}( {\bf r})$ where ${{\bf k}}$ is the wave vector (first Brillouin 
zone of the supercell) and $\nu$ is the band index. 
Subsequently, the MD step is performed by moving the ions according to the forces 
imposed by the electron system using the Hellmann-Feynman theorem. This procedure is 
repeatedly performed until thermodynamic equilibrium is reached. 
Then physical observables such as the equation of state (pressure, internal energy), 
pair distribution functions, and diffusion coefficients can be extracted. In this way,
the ion dynamics is treated properly, allowing to resolve even the collective ion 
acoustic modes~\cite{Rueter13,White13}. Furthermore, an evaluation of the Kubo formula 
is possible for a number of snapshots of the DFT-MD simulation; for details, 
see~\cite{Desjarlais02,Mazevet05,holst08,Holst11,Mattsson05,French14}.

The DFT-MD method works very well for fairly high density or coupling parameters, 
but the limiting case $\Gamma\ll 1$ and $\Theta>1$ has been addressed too, 
see~\cite{Lambert11, Wang13}. It is still an open question to what extent the $e-e$ 
correlations in the XC functional represent $e-e$ collisions in this limit as 
discussed above. The numerical results indicate that at least parts of the $e-e$ 
contributions are included.

Starting point for the calculation of the conductivity in the DFT-MD method is the Kubo 
formula (\ref{Kubo}). The equilibrium statistical operator $\hat \rho_0$ contains the Kohn-Sham 
Hamilton operator $\hat H_{\rm KS}$. The time-dependence of the operators within the Heisenberg 
picture in the correlation functions (\ref{acf}) is treated as
$
{\bf \hat P}(t-i \hbar \tau) =\e^{\frac{i}{\hbar}(t-i \hbar \tau)  \hat H_{\rm KS}} {\bf \hat P}
\e^{-\frac{i}{\hbar}(t-i \hbar \tau)  \hat H_{\rm KS}}
$. 
Single-electron states ($\hat H_{KS} |k \nu \rangle =E_{k\nu} |k \nu \rangle$) are introduced solving the Schr\"odinger 
equation for a given ion configuration within the KS approach. 
With the momentum operator (\ref{moments}) in second quantization 
$
{\bf \hat P}= \sum_{{\bf kk'}\nu\nu'} \langle {\bf k} \nu| {\bf \hat p}| {\bf k'} \nu' \rangle
  \hat a^\dagger_{{\bf k} \nu}  \hat a^{}_{{\bf k}' \nu'}
$, 
the averages with the equilibrium statistical operator are 
evaluated using Wick's theorem. From the Kubo formula (\ref{Kubo}), we find for the real part of 
conductivity
\begin{eqnarray}\label{KG1}
 {\rm Re}\, 
 \sigma^{\rm KG}_{\alpha \beta}(\omega)&=&\frac{2 \pi e^2}{3 \Omega_{c} m^2 \omega} \sum_{{\bf k}\nu\nu'}
 \langle {\bf k} \nu| {\bf \hat p}| {\bf k} \nu' \rangle \cdot \langle {\bf k} \nu'| {\bf \hat p}| {\bf k} \nu \rangle
 (f_{ {\bf k} \nu}-f_{ {\bf k} \nu'}) \delta_\eta (E_{ {\bf k} \nu}-E_{ {\bf k} \nu'}-\hbar \omega) \,.
\end{eqnarray}
Here, a broadened $\delta$ function
\begin{equation}\label{delta}
 \delta_\eta(x) = \frac{1}{\pi}\frac{\eta}{x^2+\eta^2} 
\end{equation}
is introduced and the matrix elements are given by
$ 
\langle {\bf k} \nu| {\bf \hat p}| {\bf k'} \nu' \rangle=\delta_{\bf k ,k'}\left[\hbar {\bf k} \delta_{\nu, \nu'}
+\frac{1}{\Omega_c} \int_{\Omega_c} d^3 {\bf r} u^*_{ {\bf k} \nu}( {\bf r})  (\hbar/i) (\partial /\partial {\bf \hat r}) u_{ {\bf k} \nu'}( {\bf r}) \right]
$. 

Extensive DFT-MD simulations have been performed, for instance, for warm dense hydrogen~\cite{Holst11,holst08,French12} 
using up to $N_c=512$ atoms in a supercell (depending on the density) and periodic boundary conditions
so that $N_c$ discrete bands appear in the electronic structure calculation for the cubic supercell. 
Expression (\ref{KG1})
has been evaluated numerically, where $f_{ {\bf k} \nu}= f(E_{ {\bf k} \nu})$ describes the occupation of the 
$\nu$th band, which corresponds to the energy $E_{ {\bf k} \nu}$ 
at $\bf k$.
Since a discrete energy spectrum results from the finite simulation volume $\Omega_c$, 
the $\delta_\eta$ function has to be broadened, see App. \ref{subsdeltaeta},  at least by about the minimal discrete energy difference.
An integration over the Brillouin zone is performed by sampling special $\bf k$ points, 
with a
respective weighting factor $W(\bf k)$ \cite{holst08,Desjarlais02}.
The imaginary part of the conductivity can be calculated using the Kramers-Kronig relation. 

The Kubo-Greenwood formula (\ref{KG1}) takes adequately into account $e-i$ collisions via the interaction potential 
as well as the ion-ion correlations via a structure factor.
This way to treat the $e-i$ interaction makes the transition from WDM to solid state band structure calculations more consistent.
Pseudo-potentials and ionic structure factors are correctly treated.
The $e-e$ interaction is considered in the 
KS Hamiltonian via the XC functional.
Using the representation by Bloch states $| {\bf k} \nu' \rangle$ which diagonalize the
KS Hamiltonian, the time dependence in the current-current correlation function (\ref{KG1}) 
is trivial leading to the $\delta_\eta$ function. As shown in App.~\ref{subsdeltaeta}, 
convergent results in the static case can be obtained  due to the broadening of the 
$\delta_\eta$ function (\ref{delta}). It is not clear until now whether $e-e$ collisions 
are rigorously reproduced in this approach, and more detailed investigations to solve this 
problem are planned for the future.

\section{Results}
\subsection{The correction factor for arbitrary degeneracy}\label{correction}

After discussing the correction factor (\ref{eq:Ree}) in the limit of non-degenerate hydrogen-like plasmas, we give now results for the static case $R_{\textrm{ee}}(\omega = 0)=R_{\rm ee} $ for arbitrary degeneracy that is relevant for WDM. Using the definition of the renormalization functions $r$ in the Drude-like expression (\ref{Drude2}), we can express the static correction factor as
\begin{equation}\label{reduction}
 R_{\rm ee} = \frac{\sigma_{\rm ei+ee}}{\sigma_{\rm ei}} = \frac{r_{\rm ei}}{r_{\rm ei+ee}}\,,
\end{equation}
where $r_{\rm ei+ee}=r^{(2)}(0)$ shall be calculated according to Eq.~(\ref{3sigma}) and for $r_{\rm ei}$ the $e-e$ 
contributions are neglected. In general, considering arbitrary degeneracy $\Theta$, the result depends on the plasma 
parameters $T, n_{\rm e}$ as well as the ion charge $Z$.
The correlation functions in Eq.~(\ref{3sigma}) were calculated in Born approximation. 
For the evaluation of the corresponding integrals, see App.~\ref{app0}. 
For easy access in any application we give an expression which was fitted to the numerical data. 
The following fit formula is valid in the temperature range of  
$T \gtrsim 10^4$~K up to temperatures where relativistic effects need to be taken into account, 
and free electron densities $n_{\rm e}\lesssim10^{24}$~cm$^{-3}$ with an error of less than 2\%, 
\begin{align}\label{eq:ReeFit}
  R_{\rm ee}(T,\Theta,Z)= 1-A(Z)+\left\{ \frac{1}{A(Z)} + \frac{1}{a\, B(Z)} \ln\left(1+ \left[ 
  \frac{e^{-\frac{B(Z)}{A(Z)}}}{C(T,Z)} \frac{3\sqrt{\pi}}{4}\Theta^{3/2}\right]^{a}\right) \right\}^{-1} + 
  G(T,Z)\cdot e^{-\frac{\left[ \ln(\Theta) - M(T) \right]^2}{2  \left[S(T)\right]^2}}\,,
\end{align}
where we introduced the fit coefficient $a=0.76$ and the functions 
\begin{align}
 A(Z)&=\frac{9\sqrt{2}}{13(Z+\sqrt{2})} , \label{eq:class1} \\
 B(Z)&=\frac{3 \left[ \sqrt{2} Z \left(67 + 39 \ln(2)\right) + 56\right]}
 {[13(Z+\sqrt{2})]^2} \approx \frac{21(19 Z + 8)}{[13(Z+\sqrt{2})]^2} , \label{eq:class2} \\
 C(T,Z) &= e^{\gamma}\frac{(1+Z)}{2\pi \hbar} \frac{e^2}{4\pi \epsilon_0}\sqrt{\frac{2m}{k_BT}}\approx 225.2 
 \frac{1+Z}{\sqrt{T[{\rm K}]}} , \label{eq:class3} \\
 G(T,Z) &= \frac{1}{Z} \left[ \Big( 0.0443  \ln\big( T[{\rm K}] \big) \Big)^3 - 
 \Big( 0.0476 \ln\big( T[{\rm K}] \big) \Big)^2 + 0.0185 \ln\big( T[{\rm K}] \big) - 0.0170 \right] , \label{eq:Gdeg1}\\
 M(T) &= 5.9 - 2.5 \ln\Big( \ln\big( T[{\rm K}] \big) \Big) , \label{eq:Gdeg2} \\
 S(T) &= 1 + 0.015 \ln\big( T[{\rm K}] \big) , \label{eq:Gdeg3}
\end{align}
where $\gamma$ is again Euler's constant and $\Theta$ is defined in Eq. (\ref{theta}).  
Instead of the density $n_{\rm e}$ we use the electron degeneracy parameter $\Theta$ in Eq.~(\ref{eq:ReeFit}) 
that was designed using the known limiting cases as discussed in Subsec.~\ref{subsec:limits}. 
The explicit dependence on temperature $T$ and effective charge $Z$ in Eqs.~(\ref{eq:class1})-(\ref{eq:class3}) is based on the analytical result for the classical behavior, see App.~\ref{app1}, 
and the high-density limit $\lim_{\Theta \ll 1}R_{\rm ee}(T,\Theta,Z) = 1$. 
Furthermore, we use a Gaussian-like term in the fit in order to interpolate at arbitrary degeneracy parameter 
$\Theta$, with the functions given by Eqs.~(\ref{eq:Gdeg1})-(\ref{eq:Gdeg3}), see App.~\ref{app1}.
The fit is not only valid for fully ionized hydrogen but also for WDM with any effective ionization $Z$.

\begin{figure}[htp] 
\begin{center}
 \includegraphics[width=12cm]{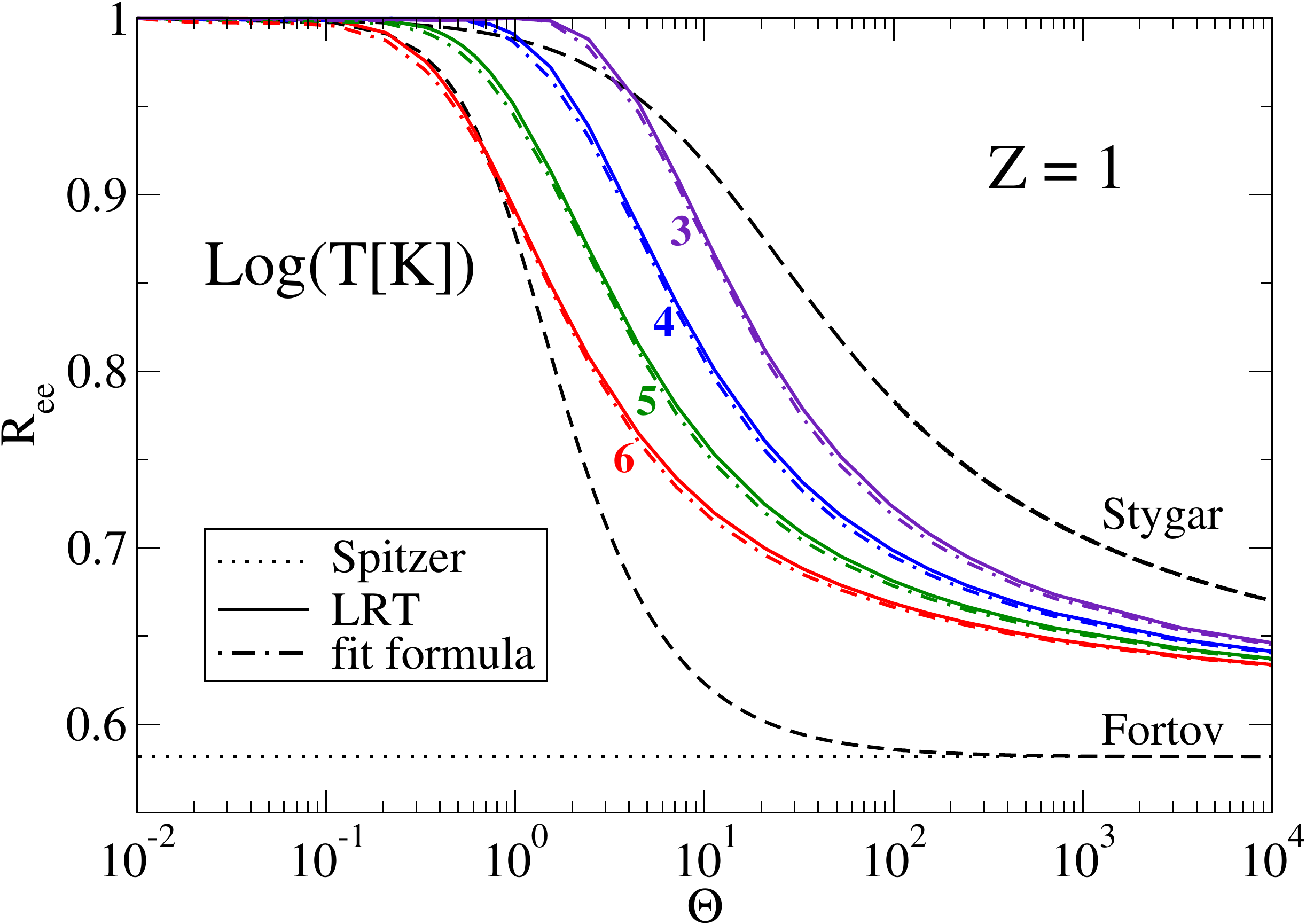}
\end{center}
  \caption{(Color online) Correction factor $R_{\rm ee}$ 
  of the conductivity due to $e-e$ collisions
  as function of degeneracy parameter $\Theta$ at $Z=1$ for different temperatures $T=(10^3, 10^4, 10^5, 10^6)$~K. 
  Numerical calculations (LRT, full lines) are compared with the fit formula (\ref{eq:ReeFit}) (dot-dashed lines) 
  and the approximations (\ref{eq:Stygar}) of 
  Stygar \textit{et al.}~\cite{stygar02} and (\ref{eq:Fortov}) of Fortov~\textit{et al.}~\cite{Fortov03} (dashed lines).}
\label{fig:TZ1}
\end{figure}

Fig.~\ref{fig:TZ1} shows the results for $Z=1$ in dependence on the density and temperature. The $e-e$ interaction generally leads to a reduction of the static conductivity which is expected due to an additional scattering process. Also, this becomes less relevant with increasing degeneracy due to the Pauli exclusion effect. Fig.~\ref{fig:TZ2} in App.~\ref{app1} illustrates the results for $Z=2$ and $3$, respectively. 
With increasing effective charge, the $e-e$ correction factor becomes smaller. 

Beside the comparison of the fit formula~(\ref{eq:ReeFit}) with the numerical results, 
Figs.~\ref{fig:TZ1} and \ref{fig:TZ2} 
show the low-density limit given by  Spitzer, 
see Eq.~(\ref{eq:ReeKT}), which would be reached at very large values of $\Theta$ only. 
Also shown are approximations proposed by Stygar \textit{et al.}~\cite{stygar02}
\begin{align}\label{eq:Stygar}
 R_{\rm ee}^{\rm Stygar}(\Theta,Z) = R_{\rm ee}^{\rm KT}(Z) + \frac{1-R_{\rm ee}^{\rm KT}(Z)}{1+0.6 \ln\left( 1+\frac{\Theta}{20} \right)} ,
\end{align}
and Fortov \textit{et al.}~\cite{Fortov03}
\begin{align}\label{eq:Fortov}
 R_{\rm ee}^{\rm Fortov}(\Theta,Z) = R_{\rm ee}^{\rm KT}(Z) + \frac{1-R_{\rm ee}^{\rm KT}(Z)}{\sqrt{1+\Theta^2}} ,
\end{align}
with the Spitzer values $R_{\rm ee}^{\rm KT}(Z=1)=0.582$, see Eq.~(\ref{eq:ReeKT}), 
and $R_{\rm ee}^{\rm KT}(Z=2)=0.683$, see \cite{Spitzer53}. 
The value $R_{\rm ee}^{\rm KT}(Z=3)=0.778$ follows from the low-density limit $1-A(Z)$ in Eq.~(\ref{eq:ReeFit}).
The phenomenologically constructed approximations of Fortov \textit{et al.} and Stygar \textit{et al.} 
do not include an explicit dependence on $T$. The Stygar \textit{et al.} expression gives the behavior in the low-density limit qualitatively correct, whereas the behavior in the region of strong degeneracy is better described by that of Fortov \textit{et al.}~\cite{Fortov03}. A numerical analysis of the correction factor using gLRT has already been presented  by Adams \textit{et al.} in Ref. \cite{Adams07}  but no fit formula was given.

The inclusion of further effects such as dynamical screening, ion-ion structure factor, and strong 
collisions (see Refs.~\cite{RR89,RRMK89,Redmer97,Reinholz05,Karachtanov13}) requires more 
detailed investigations. However, these effects are of less relevance for the correction due to $e-e$ collisions, 
both in the high-density and low-density limit. In the latter case, corrections appear only in higher orders of the virial expansion.
Dynamical screening can be taken into account approximatively by an effective screening radius, see Refs.~\cite{Karachtanov11,Karachtanov13},  
but affects the correction factor by less than 2\%.

\subsection{The contribution of $e-e$ collisions}

The discussion on the inclusion of $e-e$ collisions in the case of DFT simulations is still ongoing. However, this is crucial when comparing different approximations as will be seen in the following Subsection. Here we want to respond to an argumentation given by Dharma-wardana in Ref.~\cite{dharma06}. Using the relaxation time approach,  the single-center T-matrix  combined with a total ion-ion structure factor derived from quantum HNC was calculated. 
Comparison with data for aluminum and gold show good coincidence in the region of 
a degenerate electron system. Here, Fermi's golden rule and the relaxation time ansatz are justified which follows from our discussion as well.

A more general discussion in Ref.~\cite{dharma06} on the role of $e-e$ interaction 
for the electrical conductivity argues that  no resistivity can be observed because 
the total current is conserved under $e-e$ interaction. This seems trivial. However, 
it is not stringent to conclude that this is also the case in the general case of a 
two-component plasma.
There is an indirect influence via the screening of the electron-ion pseudo-potential 
interaction that arises within a mean-field treatment. 
Even more, those collisions are entropy producing. The umklapp processes in crystalline 
solids~\cite{ChrisRoep85} are not relevant in a plasma since there is no long-range order. 
It is correct that the interaction with the ion subsystem is necessary to obtain any change 
in the total electron current, but it cannot be said that $e-e$ interactions play no part 
in the static or dynamic conductivity at all. 

The Spitzer result takes into account the contribution of $e-e$ collisions to the conductivity. 
This is due to the flexibility of the single-momentum distribution $f(\bf p)$ that is sensitive to the 
contribution of  $e-e$ collisions. The same is also obtained introducing moments of the 
distribution function as done in the variational approach~\cite{Redmer97,Reinholz05}. It is claimed and generally accepted that 
the Spitzer result is the benchmark for the low-density limit of a classical plasma. In contrast, the conclusion drawn 
in~\cite{dharma06}, that this does not establish the validity of results of the Spitzer  
type, is not convincingly justified. The other  main argument is, that good agreement between experimental data and 
calculations neglecting $e-e$ contributions shows that the direct role of $e-e$ interactions, 
taken for granted in the plasma literature, needs to be seriously reconsidered. We have shown that it is the particular case 
of highly degenerate WDM states where the contribution of $e-e$ collisions to the conductivity becomes 
small indeed. This can readily be seen from the correction factor $R_{\rm ee}(\Theta)$ that approaches the value $1$ 
for $\Theta \ll 1$.

\subsection{Conductivity of aluminum plasma}
\label{sec:aluminium}

\begin{figure}[htp] 
\begin{center}
\includegraphics[width=10cm]{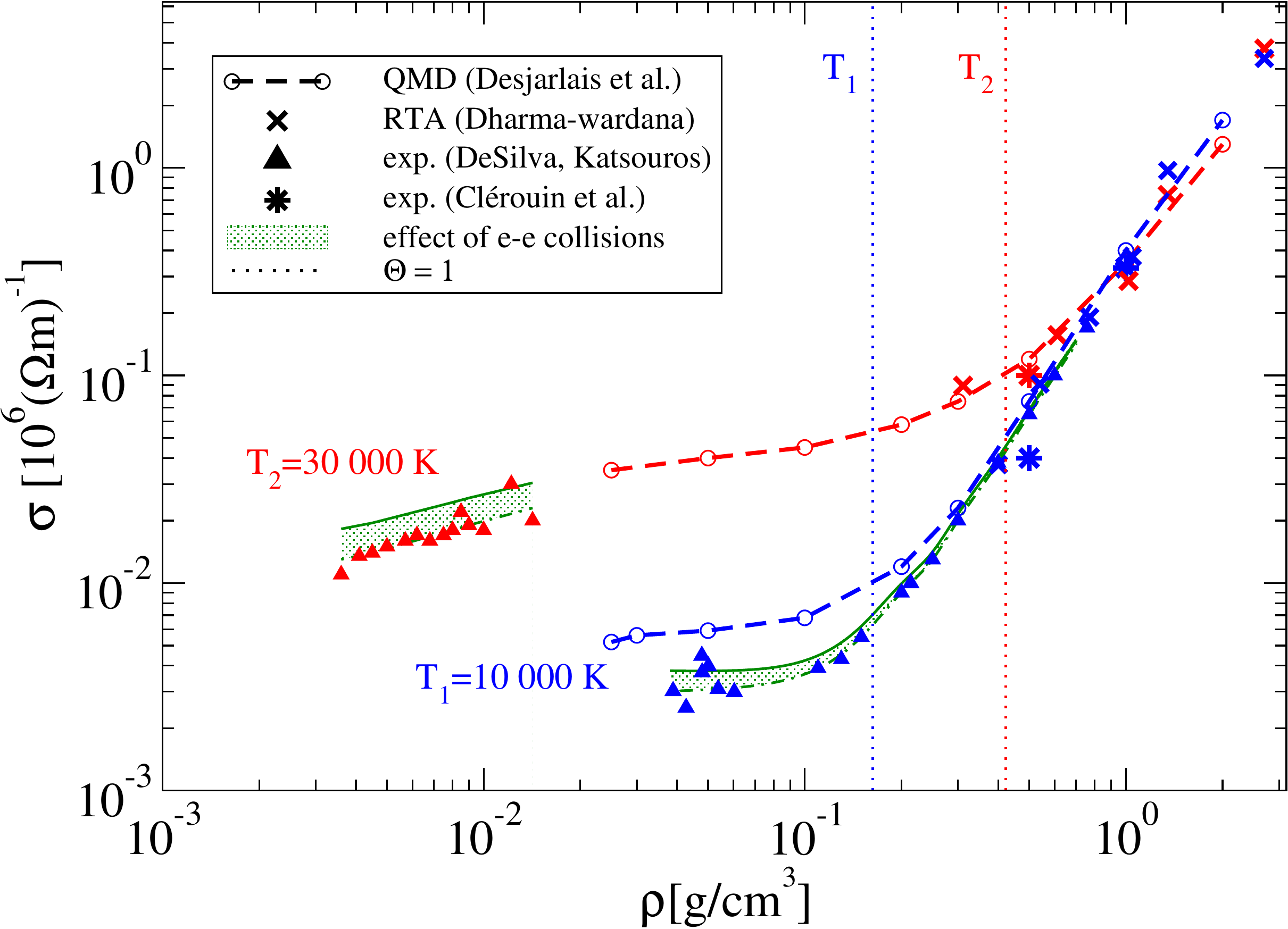}\,\,\,\,
\end{center}
  \caption{(Color online) Aluminum dc conductivity as function of density for 10~000~K (blue) and 30~000~K (red). 
  Experiments were performed by DeSilva and Katsouros~\cite{Silva} (triangles) 
  and Clerouin \textit{et al.}~\cite{Clerouin} (stars)  for which  regression curves are given by  dashed-dotted lines (green).
  The DFT-MD results of Desjarlais \textit{et al.}~\cite{Desjarlais02} are shown as hollow circles on dashed lines. 
  Calculations of Dharma-wardana~\cite{dharma06} based on the relaxation time approximation (RTA) are given as crosses.
  Degeneracy effects become important right to the vertical dotted lines ($\Theta=1$).
Solid  lines (green) show
 the conductivity of a hypothetical Lorentz plasma,  obtained by extracting the $e-i$ scattering contributions from the regression curve  (dashed-dotted line)
  according to the correction factor $R_{\rm ee}$, Eq.~(\ref{eq:ReeFit}), for the given densities and temperatures.
}
\label{fig_2b}
\end{figure}

The static electrical conductivity of Al plasma has been investigated experimentally by a number of groups, see e.g.~\cite{Silva,Clerouin,Krisch98}, and also been discussed in the context of theoretical approaches, see Kuhlbrodt \textit{et al.}~\cite{Redmer99,Kuhlbrodt00,Kuhlbrodt01} and references there in, and Refs.~\cite{dharma06,Desjarlais02,Mazevet05,Clerouin}.
Exemplarily, we consider experimental data that were theoretically analyzed by Desjarlais \etal~\cite{Desjarlais02}, see also~\cite{Mazevet05}, using the Kubo-Greenwood formula (\ref{KG1}). The results for the dc conductivity are shown in Fig.~\ref{fig_2b}. The dotted lines indicate the density for which $\Theta=1$, i.e.\ degeneracy effects are important 
to the right of these lines. 

For solid state densities, the electron system is degenerate ($\Theta<1$, region right to the dotted lines) 
and the correction factor is $R_{\rm ee} \approx 1$ there, see Figs.~\ref{fig:TZ1} and \ref{fig:TZ2}. 
The conductivity is essentially determined by the $e-i$ interaction whereas the $e-e$ interaction 
does not give a direct contribution but influences the $e-i$ pseudo-potential due to screening and exchange interactions. 
In this region, excellent agreement between the measured data~\cite{Silva,Clerouin} and the DFT-MD simulations using the 
Kubo-Greenwood approach~\cite{Desjarlais02} can be stated. Evaluations based on gLRT yield also the correct qualitative 
behavior in this region but depend on the choice for the screening function and the ion-ion structure factor~\cite{Redmer99,Kuhlbrodt00,Kuhlbrodt01}, see Eq.~(\ref{Ziman}).

At low densities, the aluminum plasma is at conditions where $\Theta>1$ so that the plasma is no longer degenerate. In this region, the correction factor is $R_{\rm ee} < 1$, see Figs.~\ref{fig:TZ1} and \ref{fig:TZ2}, so that $e-e$ collisions contribute 
to the conductivity. In order to illustrate the influence of $e-e$ collisions, we propose the following procedure. 
Dividing the measured values by the correction factor $R_{\rm ee}$, Eq.~(\ref{eq:ReeFit}), yields the contribution of the $e-i$ collisions to the conductivity, thus giving simultaneously an estimate for the effect of $e-e$ collisions. 
To apply the correction factor the charge state $Z$ has to be specified. We use the ionization degree calculated from coupled mass action laws, see Refs. [55, 58].
For the temperature of 30 000 K, at the  densities considered here a value $Z \approx 1$ has been given.
It was also found that at 10 000 K the ionization degree is much lower in this low-density region.
The calculated average charge state of $Z\approx 0.1$ indicates that at most 1/10 of the Al atoms are ionized, and correspondingly the free electron density 
$n_{\rm e}=n_{\rm ion}\approx n_{\rm atom}/10$ is also reduced.
Besides the reduced number of charge carriers, an additional scattering contribution on the neutral atoms leads to a further reduction of the the electrical conductivity as was shown in Refs.~\cite{Redmer99,Kuhlbrodt00,Kuhlbrodt01}. Within a partially ionized system this may become  the stronger effect than that of $e-e$ collisions. This might well justify taking the $e-e$ contribution into account via the correction factor instead of an explicit numerical calculation.

Please note that the electrical conductivity in this partially ionized, non-degenerate region $\Theta>0$ and $\Gamma<1$ strongly depends on the ionization degree of the plasma and the effective interaction between the electrons, ions, and neutral atoms. The calculation of corresponding mass action laws and two-particle potentials is the main problem in this region which has been addressed in chemical models, see~\cite{Redmer99,Kuhlbrodt00,Kuhlbrodt01}. Applying DFT-MD simulations in this low-density region is a challenge since most of the DFT codes are based on plane-wave expansions which become computationally expensive there. Furthermore, the XC functional has to be chosen such that the correct band-gap (ionization energy) is reproduced. Standard XC functionals, such as given by Perdew {\it et al.}~\cite{PBE}, underestimate the band gap systematically~\cite{MPD-CPP} so that, e.g., hybrid functionals~\cite{HSE} have to be applied. These issues are subject of future work.

\section{Conclusions}

We conclude that $e-e$ collisions have to be included in the low-density, 
nondegenerate region of WDM. Compared with calculations of the dc conductivity that 
take into account only $e-i$ collisions, such as the use of the 
relaxation time ansatz, the contribution of the $e-e$ collisions can be represented 
by a correction factor $R_{\rm ee}$ 
that depends mainly on the degeneracy parameter $\Theta$. 
In the case of a strongly degenerate electron gas 
($\Theta \ll 1$), the contribution of $e-e$ collisions can be neglected
since only umklapp processes are of relevance in solids. In the non-degenerate limit 
$\Theta \gg 1$, the $e-e$ collisions lead to a reduction of the dc conductivity 
by a factor of about 0.5 for $Z=1$. With increasing $Z$ the reduction becomes less relevant
leading to the Lorentz plasma result for $Z\gg 1$.

The generalized linear response theory allows to evaluate the transport coefficients 
of WDM in a wide region, joining the limits of strong degeneracy known from liquid metals
and of low densities as known from standard plasma physics. The present work considers 
free electrons interacting with ions having an effective charge $Z$.
The fit formula given in Sec.~\ref{correction} to calculate the influence of $e-e$ collisions 
on the conductivity allow for a better implementation in codes and other applications.

Future work will be concerned with the frequency dependence of the correction factor $R_{\rm ee}$. While it was already  shown numerically in \cite{Reinholz12}, that the renormalization function is not relevant in the high frequency limit, $\lim_{(\omega_{\textrm{pl}}/\omega) \rightarrow 0} r(\omega) =1$, the intermediate frequency region has to be investigated for any degeneracy.

The implementation of pseudo-potentials and the ion-ion structure factor  become 
of relevance with increasing free electron density. However, at high densities, the influence of the 
renormalization function is fading, $r(\omega) \to 1$. Therefore these effects are of high relevance for the 
$e-i$ collisions determining the collision frequency $\nu^{\textrm{Ziman}}$, but barely  relevant for 
the correction factor $R_{\rm ee}$. 

Another issue is the composition of WDM in the low-density, low-temperature limit where a chemical 
model is applicable. The ionization degree and composition are derived from a mass action law, 
that gives the effective charge $Z$ in dependence of temperature $T$ and ion density $n_{\rm ion}$.
In particular, for the partially ionized plasma, additional scattering with neutrals will reduce the 
conductivity at low temperatures considerably. Further work is necessary in order to relate
predictions of chemical models to those based on DFT, and to clarify the role of $e-e$ collisions 
within DFT-MD in the low-density limit.

\begin{acknowledgments}
We thank J.\ Adams, J.\ Cl\'erouin, M.P.\ Desjarlais, M.W.C.\ Dharma-wardana, M.\ French, 
and V.S.\ Karakhtanov for fruitful discussions of problems presented in this paper. 
The authors acknowledge support from the DFG within the Collobarative Research Center SFB~652.  
\end{acknowledgments}

\appendix

\section{Generalized linear response theory}
 \label{app:gLRT}

In the case of a charged particle system considered here, described by the Hamiltonian $\hat H$, under the influence of an external field, $\hat H_{F}(t)=-e {\bf \hat r}{\bf E}^{\rm ext} (t)$, 
the non-equilibrium statistical operator has to be determined. 
Following Zubarev \cite{ZMR1,Roep98,Reinholz05,Reinholz00}, one  starts with a relevant statistical operator 
\begin{equation} \label{rhorel}
\hat \rho_{\rm rel}(t)=\frac{1}{Z_{\rm rel}(t)}\e^{-\beta(\hat H-\mu \hat N) +  \sum_l F_l(t) \hat B_l}, \qquad
Z_{\rm rel}(t)= \tr {\e^{-\beta(\hat H-\mu \hat N) +  \sum_l F_l(t) \hat B_l}}\,,
\end{equation}
as a generalized Gibbs ensemble which is derived from the principle of maximum of the entropy.
This relevant distribution is characterized by a set of relevant observables $\{\hat B_l\}$ chosen in addition to energy $\hat H$ and number of particles $\hat N$.
The Lagrange parameters $\beta, \mu, F_l(t)$, which are real valued numbers, are introduced to fix the given averages
\begin{equation}
\label{selfc}
\tr{\hat B_l \, \hat \rho(t)}=\langle \hat B_l\rangle^t =\tr { \hat B_l\,\hat \rho_{\rm rel}(t)}\,.
\end{equation}
These  self-consistent  conditions  ensures that the observed averages $\langle  \hat B_l\rangle^t $ are exactly reproduced 
by the hermitian $ \hat \rho_{\rm rel}(t)$. Similar relations are used in equilibrium to eliminate the Lagrange parameters $\beta$ and $\mu$. 
Starting with the relevant statistical operator, the stationary non-equilibrium   state is formed dynamically, and this 
process converges the faster the more relevant observables $\hat B_l$ are included to characterize the initial state.
The selection of the set of relevant observables has no influence on the result if the calculations are performed rigorously,
but will influence the result if approximations such as perturbation expansions are performed.

In linear response, the response parameters  $F_l(t)$ are considered to be small so that we can solve the
implicit relation (\ref{selfc}). 
The response parameter are determined after expanding up to the first order 
with respect to the external field ${\bf E}^{\text{ext}}(\omega)$ (we consider a homogeneous field, e.g. zero  wave vector) 
and the response parameters
$ F_l$, where $F_l(t)=  {\rm Re}\{ F_l \e^{-i \omega t}\}$.
We arrive at the response equations \cite{Reinholz12}
\begin{equation} \label{LBE1c}
\sum_{l'} \left[\left( { \hat B}_l ;\dot{\hat B}_{l'}\right)+\left<  \dot{\hat B}_l ; \dot{\hat B}_{l'}\right>_z - 
i \omega \left\{\left(  { \hat B}_l ;{ \hat B}_{l'}\right)+  \left<   \dot{\hat B}_l ;\delta { \hat B}_{l'} \right>_z \right\} \right] F_{l'}
=
\beta \frac{e}{m}\left\{\left(  { \hat B}_l ;{\bf \hat P}\right)+  \left<   \dot{\hat B}_l ;{ \bf \hat P} \right>_z \right\}  
\cdot {\bf E}^{\text{ext}}(\omega)\,
\end{equation}
with  $z=\omega+i \eta$ ($\lim_{\eta \to +0}$) and the Laplace transform of the correlation functions, Eq. (\ref{acf}). 
The time derivative of the position operator in $ \hat H_{F}(t)$  leads to the total momentum ${\bf \hat P}=\sum_p {\bf p} \, \hat n_p$, and subsequently to the right hand side of \rf{LBE1c}. 

Considering $L$ relevant observables $  \delta \hat B_l =\hat B_l-   {\rm Tr}\{\hat B_l \rho_0 \}$,  \rf{LBE1c} is  a system of $L$ linear equations 
to determine the response parameters  $F_l$ for a given external field ${\bf E}^{\text{ext}}(\omega)$. It is the most general 
form of LRT, allowing for arbitrary choice of relevant observables $ \hat B_l$ and corresponding response parameters $F_l$.
Comparing with kinetic theory  \cite{Reinholz12}, the first correlation function  on the left hand side  can be identified as a collision term, 
while the right hand side represents the drift term due to the external perturbing field. 

The set of relevant observables $ \hat B_l$ to characterize the non-equilibrium state can be chosen arbitrarily, 
and the calculated non-equilibrium properties are independent on this choice provided
 no approximations like perturbation expansions are performed. At least, the set of relevant observables $\hat B_l$
should contain conserved quantities that determine the equilibrium state. Conveniently, also  long-living 
fluctuations in the system that are hardly produced by the dynamical evolution (such as bound state formation) should be taken into account.
Otherwise, a perturbation expansion is  converging only slowly. 
Different expressions and results can be understood as approximations, working in a Markov approximation and describing
the system on different levels of sophistication. Results that are obtained in lowest order are improved summing up 
(sometimes divergent) terms that occur in higher order perturbative expansions. Alternatively, we can suggest  different 
choices of the set of relevant observables $\hat B_l$ like a variational approach (Kohler variational principle),
see \cite{Reinholz12}.

Starting with the occupation numbers $\hat n_p$ of the single-particle states $|p\rangle$ as set of relevant observables $ \hat B_l$, 
we arrive at the generalized linear Boltzmann equations \cite{Reinholz12}
($\delta \dot {\hat n}_{p}= \dot {\hat n}_{p}$)
\begin{equation} \label{LBE}
\sum_{p'} \left[(\delta  \hat n_p,\dot {\hat n}_{p'})+\left< 
\dot {\hat n}_{p}; \dot {\hat n}_{p'} \right>_z - i \omega \left\{ (\delta {\hat n}_{p},\delta \hat n_{p'}) + 
\left<  \dot {\hat n}_{p}; \delta {\hat n}_{p'}\right>_z \right\}\right] F_{p'}
=
\frac{e  }{m}\beta\sum_{p''}
\left[  (\delta  \hat n_p, {\hat n}_{p''}) + \left< \dot {\hat n}_{p};  {\hat n}_{p''}\right>_z \right] {\bf p}''  
\cdot {\bf E}^{\text{ext}}(\omega)\,.
\end{equation}
This is the basic equation to work out the linear response approach given in Sec. \ref{fluctuations}.

\section{Calculation and simplification of the correlation functions, Eqs.~(\ref{deiLB}, \ref{deeLB})}
\label{app0}
The expression (\ref{deiLB}),
\begin{eqnarray}
\label{deiLBa}
d_{ll'}^{\rm ei}&=& \pi\hbar Z^2 2 (2s_i+1)\sum_{\bf{k}\bf{p}\bf{q}}\int\limits_{-\infty}^{\infty}d \hbar \omega
\left|\frac{V(q)}{\epsilon^{\rm RPA}(q,\omega)}\right|^{2}f_{k}^{e}(1-f_{\left|\bf{k}+\bf{q}\right|}^{e}) 
f_{p}^{i}(1-f_{\left|\bf{p}-\bf{q}\right|}^{i})\nonumber \\ 
 &&\times \delta(\hbar \omega -E_{\left|\bf{k}+\bf{q}\right|}^{e}+E_{k}^{e})\delta(\hbar \omega 
 -E_{p}^{i}+E_{\left|\bf{p}-\bf{q}\right|}^{i}) 
 K_{l} ({\bf k},{\bf q}) K_{l'}({\bf k}, {\bf q}) \,,
\end{eqnarray} 
is evaluated by performing the integral over $\hbar \omega$. In the resulting $\delta$ function that describes
energy conservation, we can neglect the ionic contributions because of the large mass ratio (adiabatic limit, 
elastic collisions of the electrons at the fixed ions). The ions are treated classically, and the summation over $\bf p$ 
and spin summation gives simply 
$(2 s_i+1)\sum_{\bf p} f^i_p = n_{\rm ion} \Omega=N_{\rm ion}$, the number of ions.
In particular,
\begin{eqnarray}
\label{deiLBb}
d_{11}^{\rm ei}&=&2 \pi\hbar Z^2 n_{\rm ion} \Omega\sum_{\bf{k}\bf{q}}
\left[\frac{e^2}{\epsilon_0 \Omega (q^2+\kappa^2)}\right]^{2}f_{k}^{e}(1-f_{\left|\bf{k}+\bf{q}\right|}^{e}) 
 \delta(E_{k}^{e}-E_{\left|\bf{k}+\bf{q}\right|}^{e})q_z^2 \,,
\end{eqnarray} 
or with $q_z^2 \to q^2/3$ and transforming the $\delta$ function
\begin{eqnarray}
\label{deiLBc}
d_{11}^{\rm ei}&=& \frac{e^4}{(4 \pi \epsilon_0)^2}\frac{32 \pi^2}{3} \pi\hbar Z^2 n_{\rm ion} \Omega \frac{4 \pi \,2 \pi}{(2 \pi)^6}
\int_0^\infty dq \frac{q^4}{(q^2+\kappa^2)^2} \int_0^\infty dk \,k^2 \, f_k^e (1-f_k^e) \int_{-1}^1 dz 
\,\delta\left(z+\frac{q}{2 k}\right) 
\frac{m}{\hbar^2 k q}\,.
\end{eqnarray} 
Now, the integral over $z$ can be performed so that $k \le q/2$, 
and we transform the $k$ integral as $k\, dk = dk^2/2=d(\beta E_k) m/(\hbar^2 \beta)$ 
(note that the superscript $e$ for electrons is omitted throughout the rest of this appendix),
\begin{eqnarray}
\label{deiLBd}
d_{11}^{\rm ei}&=& \frac{e^4}{(4 \pi \epsilon_0)^2}\frac{4}{3 \pi^2} \pi\hbar Z^2 n_{\rm ion} \Omega \frac{m}{\hbar^2}
\int_0^\infty dq \frac{q^3}{(q^2+\kappa^2)^2}\left(-\frac{m}{\hbar^2 \beta}\right)\int_{\beta \hbar^2 q^2/(8m)}^\infty 
d(\beta E_k) \frac{d}{d(\beta E_k)}\frac{1}{e^{\beta E_k- \beta \mu_e^{\rm id}}+1} \,,
\end{eqnarray} 
so that the integral over $k$ is performed,
\begin{eqnarray}
\label{deiLBe}
d_{11}^{\rm ei}&=& \frac{e^4}{(4 \pi \epsilon_0)^2}\frac{4}{3 \pi^2} \pi\hbar Z^2 n_{\rm ion} \Omega \frac{m^2}{\hbar^4 \beta}
\int_0^\infty dq \frac{q^3}{(q^2+\kappa^2)^2}\frac{1}{e^{\beta \hbar^2 q^2/(8 m)- \beta\mu_e^{\rm id}}+1}
\end{eqnarray} 
or, using dimensionless $Q=\sqrt{\beta \hbar^2 q^2/m}$,
\begin{eqnarray}
\label{deiLBf}
d_{11}^{\rm ei}&=& Zd \frac{2}{n_{\rm e} \Lambda_e^3}
\int_0^\infty dQ \; \frac{Q^3}{(Q^2+\frac{\hbar^2 \beta}{m} \kappa^2)^2} \;\frac{1}{e^{Q^2/8-\alpha}+1} \,,
\end{eqnarray} 
with  $\alpha = \beta \mu_e^{\rm id}$, the thermal wavelength $\Lambda_e$ (see below Eq.~(\ref{deeLB}) in Sec.~\ref{sec:r}) and the prefactor 
\begin{equation}
 d=\frac{4}{3} (2\pi)^{1/2} Z^2n_{\rm ion}^2 \Omega m^{1/2} \beta^{1/2} \frac{e^4}{(4\pi\epsilon_0)^2}. 
\end{equation}

In analogy to Eq.~(\ref{deiLBf}) we now calculate the correlation functions (\ref{deiLBa}) with  higher moments.  
With $K_1({\bf k},{\bf q})=-q_z, \,\,\, K_3({\bf k},{\bf q})=-q_z (\beta E_k)$, 
and replacing $(\beta E_k) \to x$  we find
\begin{eqnarray}
\label{deiLBg}
d_{ll'}^{\rm ei}&=& Zd \frac{2}{n_{\rm e} \Lambda_e^3}
\int_0^\infty dQ \frac{Q^3}{(Q^2+ \frac{\hbar^2 \beta}{m} \kappa^2)^2} S_{ll'}(Q) ,
\end{eqnarray} 
with
$$S_{11}(Q)=\frac{1}{e^{Q^2/8-\alpha}+1} , $$

$$S_{13}(Q)=S_{31}(Q)=\frac{Q^2/8}{e^{Q^2/8-\alpha}+1}+\int_{Q^2/8}^\infty dx \frac{1}{e^{x-\alpha}+1} , $$

$$S_{33}(Q)=\frac{Q^4/64}{e^{Q^2/8-\alpha}+1}+2\int_{Q^2/8}^\infty dx \frac{x}{e^{x-\alpha}+1} . $$

We evaluate  the $e-e$ correlation functions, Eq.~(\ref{deeLB}), in the lowest non-vanishing order. 
Because of total momentum conservation, $d_{11}^{\rm ee}=d_{13}^{\rm ee}=0$. 
The first and only correlation function  within two-moment approach is 
\begin{eqnarray}
\label{deeLBa}
&&d_{33}^{\rm ee}= 
2\pi\beta^{2}\hbar\sum_{\bf{k}\bf{p}\bf{q}}\int\limits_{-\infty}^{\infty}d\hbar \omega
\left|\frac{V(q)}{\epsilon^{\rm RPA}(q,\omega)}\right|^{2}
f(E_{k})(1-f(E_{k}+\hbar \omega)) f(E_{p})(1-f(E_{p}-\hbar \omega))
\\ &&\nonumber \times \delta(\hbar \omega -E_{\left|\bf{k}+\bf{q}\right|}+E_{k})
\delta(\hbar \omega -E_{p}+E_{\left|\bf{p}-\bf{q}\right|}) 
\left[k_z  E_k-(k_z+q_z)(  E_k+ \hbar \omega)+p_z  E_p-(p_z-q_z)(  E_p- \hbar \omega)\right]^2.
\end{eqnarray}  
The dynamically screened Coulomb potential will be replaced by the static Debye potential, see Sec. \ref{sec:r}. The effect of dynamical screening that leads to 
the Lenard-Balescu expression for the conductivity has been discussed elsewhere \cite{Redmer97}.
For the evaluation, using spherical coordinates, we obtain 
\begin{eqnarray}
\label{deeLBb}
d_{33}^{\rm ee}&=&
\beta^{2}\hbar \frac{2\pi\Omega^3}{ 3(2 \pi)^9}  \int d^3q\int\limits_{-\infty}^{\infty}d\hbar \omega\int d^3p\int d^3k
\left|\frac{e^2}{\epsilon_0 \Omega (q^2+\kappa^2)}\right|^{2}
f(E_{k})(1-f(E_{k}+\hbar \omega)) f(E_{p})(1-f(E_{p}-\hbar \omega))
\nonumber\\ &&\nonumber \times \delta\left(\hbar \omega -\frac{\hbar^2 kq \cos \theta_k}{m}-\frac{\hbar^2 q^2}{2m}\right) 
\delta(\hbar \omega -\frac{\hbar^2 pq \cos \theta_p}{m}+\frac{\hbar^2 q^2}{2m})
\\ && \times 
 \left[q^2 (E_p- E_k)^2+2 {\bf q}\cdot ({\bf p}-{\bf k}-2 {\bf q}) (E_p- E_k) \hbar \omega
+({\bf p}-{\bf k}-2 {\bf q})^2  \hbar^2 \omega^2\right].
\end{eqnarray}
The angles between the $q$-direction and the direction of $k$ or $p$ are denoted by $\theta_k$ and $\theta_p$, respectively.

The square brackets  written in spherical coordinates are
\begin{eqnarray}
&&\left[\frac{\hbar^4}{4 m^2 } q^2 (p^2- k^2)^2+ \frac{\hbar^2}{ m }(pq \cos \theta_p-qk \cos \theta_k-2 q^2) (p^2- k^2) \hbar \omega
\right.\\&& \left.
+(p^2+k^2+4 q^2-4 pq \cos \theta_p+4 kq \cos \theta_k - 2pk (\cos \theta_p \cos \theta_k
+\sin \theta_p \sin \theta_k (\cos \phi_p \cos \phi_k+\sin \phi_p \sin \phi_k) ) \hbar^2 \omega^2\right] . \nonumber
\end{eqnarray}  
The last parentheses can be rewritten as $\cos \phi_p \cos \phi_k+\sin \phi_p \sin \phi_k=\cos(\phi_p-\phi_k)$. $\phi_p-\phi_k$ can be introduced as 
new variable, the integral vanishes.
We are left with 
\begin{eqnarray}
\label{deeLBc}
&&d_{33}^{\rm ee}= 
\beta^{2}\hbar \frac{4\pi\Omega}{ 3(2 \pi)^6} \frac{m^2}{\hbar^4}
\int_0^\infty dq\, q^2\int\limits_{-\infty}^{\infty}d\hbar \omega\int_0^\infty dk\,k^2 \int_{-1}^1 dz_k \int_0^\infty dp\,p^2\int_{-1}^1 dz_p
\,\frac{e^4}{ (q^2+\kappa^2)^2}
\\ &&\nonumber \times
f(E_{k})(1-f(E_{k}+\hbar \omega)) f(E_{p})(1-f(E_{p}-\hbar \omega))
 \frac{1}{ kq}\delta\left(z_k+\frac{ q}{2k}-\frac{m \omega}{\hbar kq} \right)
\frac{1}{ pq}\delta\left(z_p-\frac{ q}{2p}-\frac{m \omega}{\hbar pq}\right )
\\ && \times 
\left[q^2 \frac{\hbar^4}{4 m^2}(p^2- k^2)^2+2 (q p z_p-q kz_k-2 q^2) \frac{\hbar^2}{2 m}(p^2- k^2) \hbar \omega
+(p^2+k^2+4 q^2-4 pq z_p+4 kq z_k - 2pk z_p z_k)  \hbar^2 \omega^2\right].\nonumber
\end{eqnarray}  
Introducing dimensionless variables $Q$ as defined above and $x=\sqrt{\beta E_{k}}, \; y=\sqrt{\beta E_{p}},\;
\omega = \nu Q/(\beta \hbar)$, 
and performing the integrals over $z_p,z_k$, we have
%
%
%
%
\begin{eqnarray}
\label{deeLBe}
d_{33}^{\rm ee}&=& \frac{16\pi e^4 \Omega m^{7/2}}{3 \epsilon_0^2 (2 \pi)^6\beta^{5/2}\hbar^6} 
\int_0^\infty dQ\, \frac{Q^3 }{(Q^2+ \frac{\hbar^2 \beta}{m} \kappa^2)^2} \int\limits_{-\infty}^{\infty}d\nu \int_{|\nu-Q/2|/\sqrt{2}}^\infty dx\,x
\int_{|\nu+Q/2|/\sqrt{2}}^\infty dy\,y 
\\ &&\nonumber \times
\frac{1}{e^{x^2-\alpha}+1} \frac{1}{1+e^{-x^2- \nu Q+\alpha}}\frac{1}{e^{y^2-\alpha}+1} \frac{1}{1+e^{-y^2+ \nu Q+\alpha}}\\ &&\nonumber \times
\left[ (y^2- x^2)^2-2(y^2- x^2)\nu Q+2(y^2+ x^2)\nu^2 -4\left(\frac{\nu^2}{2}-\frac{Q^2}{8}\right)  \nu^2 \right].\nonumber
\end{eqnarray}  
Now we substitute  $x^2=\hat x+\nu^2/2-\nu Q/2+Q^2/8,\,\,\,y^2=\hat y+\nu^2/2+\nu Q/2+Q^2/8$, thus shifting the lower bound of the $x$ and $y$ 
integral to zero. In general the final expression
\begin{eqnarray}
\label{deeLBg}
&d_{33}^{\rm ee}&=
\frac{d}{\sqrt{2\pi}} \frac{2}{ n_{\rm e}^2 \Lambda_e^6} \int_0^\infty dQ \frac{Q^3}{\left(Q^2+ \frac{\hbar^2 \beta}{m} \kappa^2\right)^2}  f_{33}^{\rm ee}(\alpha, Q)
 \qquad \textrm{with} \nonumber\\ \label{fee33}
& f_{33}^{\rm ee}(\alpha, Q)&=
 \int\limits_{-\infty}^{\infty}d\nu \int_{0}^\infty d{\hat x}
\int_{0}^\infty d{\hat y} \left[ ({\hat y}- {\hat x})^2+2 \nu^2({\hat y}+{\hat x}) \right]
 \frac{1}{e^{{\hat x}+\nu^2/2-\nu Q/2+Q^2/8-\alpha}+1} \\&&\times  \frac{1}{1+e^{-{\hat x}-\nu^2/2- \nu Q/2+\alpha-Q^2/8}}
\frac{1}{e^{{\hat y}+\nu^2/2+\nu Q/2+Q^2/8-\alpha}+1} \frac{1}{1+e^{-{\hat y}-\nu^2/2+ \nu Q/2+\alpha-Q^2/8}} , \nonumber
\end{eqnarray}  
is evaluated numerically. For the classical limit an analytical expression can be given, see App. \ref{app1}.

\section{Correlation functions in the classical limit and construction of a fit formula}
\label{app1}
Expressions for the correlation functions derived in App.~\ref{app0} are further analyzed in the limit of non-degeneracy. We introduce integrals of the form: 
\begin{align}
 J_{1,b} &= \int \limits_0^{\infty} dQ \frac{Q^3}{(Q^2 + \frac{\hbar^2 \beta}{m}\kappa^2)^2} \frac{1}{e^{Q^2/8 - \alpha} + 1} \left(\frac{Q^2}{8}\right)^b, \\
 J_{2,b} &= \int \limits_0^{\infty} dQ \frac{Q^3}{(Q^2 + \frac{\hbar^2 \beta}{m}\kappa^2)^2} \int\limits_{0}^{\infty} dx \frac{(x+Q^2/8)^b}{e^{x+Q^2/8-\alpha}+1}, \\
 J_3 &= \int\limits_{0}^{\infty} dQ \frac{Q^3}{(Q^2 + \frac{\hbar^2 \beta}{m}\kappa^2)^2} f_{33}^{\rm ee}(\alpha, Q),
\end{align}
For the correlation functions, Eqs.~(\ref{deiLBg}, \ref{deeLBg}), we find:
\begin{align}
 \frac{d_{11}}{\Omega}&= Zd \frac{2}{n_{\rm e} \Lambda_e^3} \cdot J_{1,b=0}, \\
 \frac{d_{13}}{\Omega}&= Zd \frac{2}{n_{\rm e} \Lambda_e^3} \cdot \left(J_{1,b=1} + J_{2,b=0}\right), \\
 \frac{d_{33}^{\rm ei}}{\Omega}&= Zd \frac{2}{n_{\rm e} \Lambda_e^3} \cdot \left(J_{1,b=2} + 2 \cdot J_{2,b=1}\right), \\
 \frac{d_{33}^{\rm ee}}{\Omega}&= \frac{d}{\sqrt{2\pi}} \frac{2}{ n_{\rm e}^2 \Lambda_e^6} \cdot J_3.
\end{align}
In the classical limit ($\alpha << 0$, $\kappa^2 \approx \kappa^2_{\rm D}=\beta (1+Z) n_{\rm e} e^2 /\epsilon_0 $) the integrals yield
\begin{align}
 J_{1,b} &= e^\alpha \int \limits_0^{\infty} dQ \frac{Q^3}{(Q^2 + \frac{\hbar^2 \beta}{m}\kappa^2)^2} e^{-Q^2/8} \left(\frac{Q^2}{8}\right)^b = e^\alpha \cdot \begin{cases} -\frac{1}{2}\mathrm{Ei}\left(-\tilde k^2\right) & b=0 \\ \frac{1}{2} & b=1; 2 \end{cases}
, \\
 J_{2,b=0} &= e^\alpha \int \limits_0^{\infty} dQ \frac{Q^3}{(Q^2 + \frac{\hbar^2 \beta}{m}\kappa^2)^2} e^{-Q^2/8} = J_{1,b=0}, \\
 J_{2,b=1} &= e^\alpha \int \limits_0^{\infty} dQ \frac{Q^3}{(Q^2 + \frac{\hbar^2 \beta}{m}\kappa^2)^2} e^{-Q^2/8}\left(\frac{Q^2}{8}+1\right) = J_{1,b=0} + J_{1,b=1}, \\
 J_3 &= e^{2\alpha} \int \limits_0^{\infty} dQ \frac{Q^3}{(Q^2 + \frac{\hbar^2 \beta}{m}\kappa^2)^2} e^{-2Q^2/8} \underbrace{\int\limits_{-\infty}^{\infty}d\nu \int\limits_{0}^{\infty}d{\hat x} \int\limits_{0}^{\infty}d{\hat y} \, e^{-{\hat x}-{\hat y}-\nu^2}[({\hat y}-{\hat x})^2+2\nu^2({\hat y}+{\hat x})]}_{4 \sqrt{\pi}} \\ &= -4 \sqrt{\pi} e^{2\alpha} \cdot \frac{1}{2}\mathrm{Ei}\left(-2 \tilde k^2\right) = -4 \sqrt{\pi} e^{2\alpha} \cdot \left(\frac{1}{2}\mathrm{Ei}\left(-\tilde k^2\right) + \frac{\ln(2)}{2}\right),
\end{align}
with the coefficient $\tilde k^2=\tilde k^2(\alpha)=\frac{\hbar^2\beta}{8m}\kappa_{\rm D}^2=\frac{(1+Z)\cdot e^2}{\pi^2\hbar\epsilon_0}e^\alpha \sqrt{\frac{2m}{k_BT}}$, the exponential integral $\mathrm{Ei}(x)=-\int\limits_{-x}^{\infty} \frac{e^{-t}}{t}dt = \gamma + \ln|x| + {\cal O}(x)$ and $\gamma$ as Euler's constant, see Sec.~\ref{subsec:limits}. The term in order of $\tilde k^2$ is neglected, we approximate $\mathrm{Ei}\left(-\tilde k^2\right)\approx \gamma + \ln\left|-\tilde k^2\right|$, and therefore $\mathrm{Ei}(-2 \tilde k^2) \approx \mathrm{Ei}(-\tilde k^2) + \ln(2)$.
We obtain for the fractions of correlation functions:
\begin{align}
 \frac{N_{13}}{N_{11}} &= \frac{5}{2}, \\
 \frac{d_{13}}{d_{11}} &= 1 - \frac{1}{\mathrm{Ei}(-\tilde k^2)}, \\
 \frac{d_{33}^{\rm ei}}{d_{11}} &= 2 - \frac{3}{\mathrm{Ei}(-\tilde k^2)}, \\
 \frac{d_{33}^{\rm ee}}{d_{11}} &= \frac{\sqrt{2}}{Z}\left(1+ \frac{\ln(2)}{\mathrm{Ei}(-\tilde k^2)} \right).
\end{align}
For the renormalization functions (\ref{3sigma}) in two-moment approximation in the classical case we find 
\begin{align}
 r_{\rm ei}^{\rm cl}(\alpha) &= \frac{4}{13} - \frac{84}{169} \frac{1}{{\rm Ei}(-\tilde k^2)} + {\cal O}\left(\frac{1}{{\rm Ei}(-\tilde k^2)}\right)^2, \\
 r_{\rm ei+ee}^{\rm cl}(\alpha) &= \frac{4 (Z+\sqrt{2})}{13Z + 4 \sqrt{2}} + \frac{12Z \left[\sqrt{2}  \left(\ln(8)+4\right)-7Z\right]}{(13Z+4 \sqrt{2})^2}\frac{1}{\mathrm{Ei}(-\tilde k^2)} + {\cal O}\left(\frac{1}{{\rm Ei}(-\tilde k^2)}\right)^2,
\end{align}
for the Lorentz plasma and the plasma with $e-e$ correlations, respectively.
The correction factor (\ref{reduction}) is then given as
\begin{align}
 R_{\rm ee}^{\rm cl}(\alpha) &= \frac{r_{\rm ei}(\alpha)}{r_{\rm ei+ee}(\alpha)} \\
  &= 1- \frac{9\sqrt{2}}{13(\sqrt{2}+Z)} - \frac{3 \left[\sqrt{2}Z \left(67 + 39 \ln(2)\right) + 56\right]}{169 (Z+\sqrt{2})^2}\frac{1}{{\rm Ei}(-\tilde k^2)} + {\cal O}\left(\frac{1}{{\rm Ei}(-\tilde k^2)}\right)^2.
\end{align}
Instead of the degeneracy $\alpha$ the correction factor can be rewritten as a function of the degeneracy parameter $\Theta$
\begin{align} \label{eq:Rclass}
 R_{\rm ee}(\Theta \gg 1) &= 1- A(Z) +B(Z) \left(\ln \frac{1}{C(T,Z)}  \frac{3\sqrt{\pi}}{4}\Theta^{3/2} \right)^{-1},
\end{align}
because of $\alpha \approx \ln(\frac{4}{3\sqrt{\pi}}\Theta^{-3/2})$ in classical regimes. 
The functions $A(Z)$, $B(Z)$ and $C(T,Z)$ are given in Eqs.~(\ref{eq:class1}-\ref{eq:class3}).

In the classical limit, the asymptotic behavior of the correction factor with respect to the temperature is given analytically 
with Eq.~(\ref{eq:Rclass}). 
In the degeneracy limit ($\alpha \gg 0$, $\Theta \ll 1$), the correlation function $d_{33}^{\rm ee}=0$, so that the correction factor $R_{\rm ee}=1$. 
Therefore we construct a fit-function $R_{\rm ee}$ in which the analytical classical result goes to $1$ for high degeneracy, 
see the first three terms of Eq.~(\ref{eq:ReeFit}) in relation to Eq.~(\ref{eq:Rclass}). Eq.~(\ref{eq:ReeFit}) includes a fit 
coefficient $a$ which doesn't affect the classical limit and can be used for a better adjustment in the intermediate range.
Finally, the discrepancy between our fit formula and numerical results was reduced by a Gaussian-like term, 
see the last term of Eq.~(\ref{eq:ReeFit}). 

\begin{figure}[htp]
\begin{center}
 \hspace*{5mm}
 \includegraphics[width=8cm]{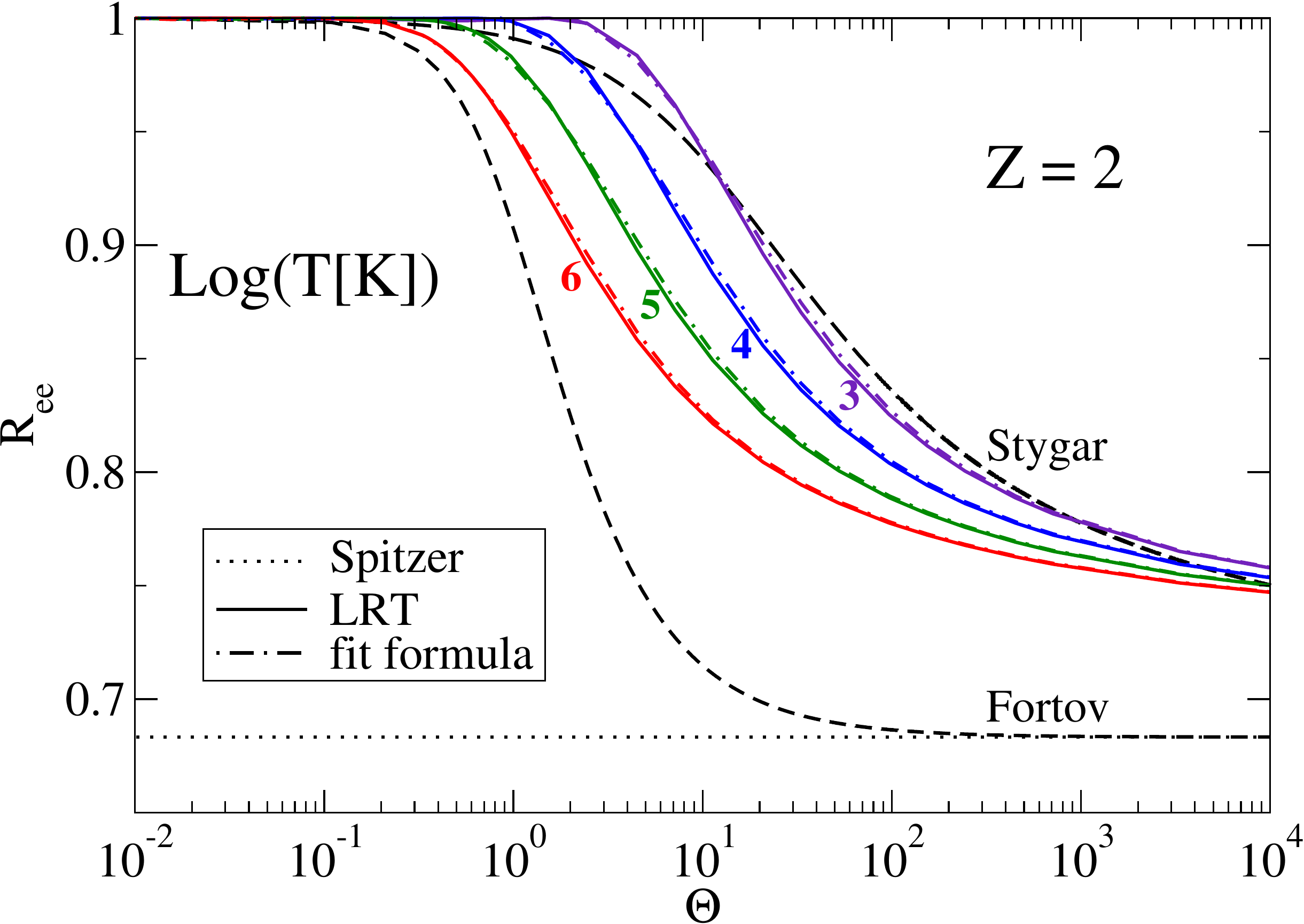}
 \hfill
 \includegraphics[width=8cm]{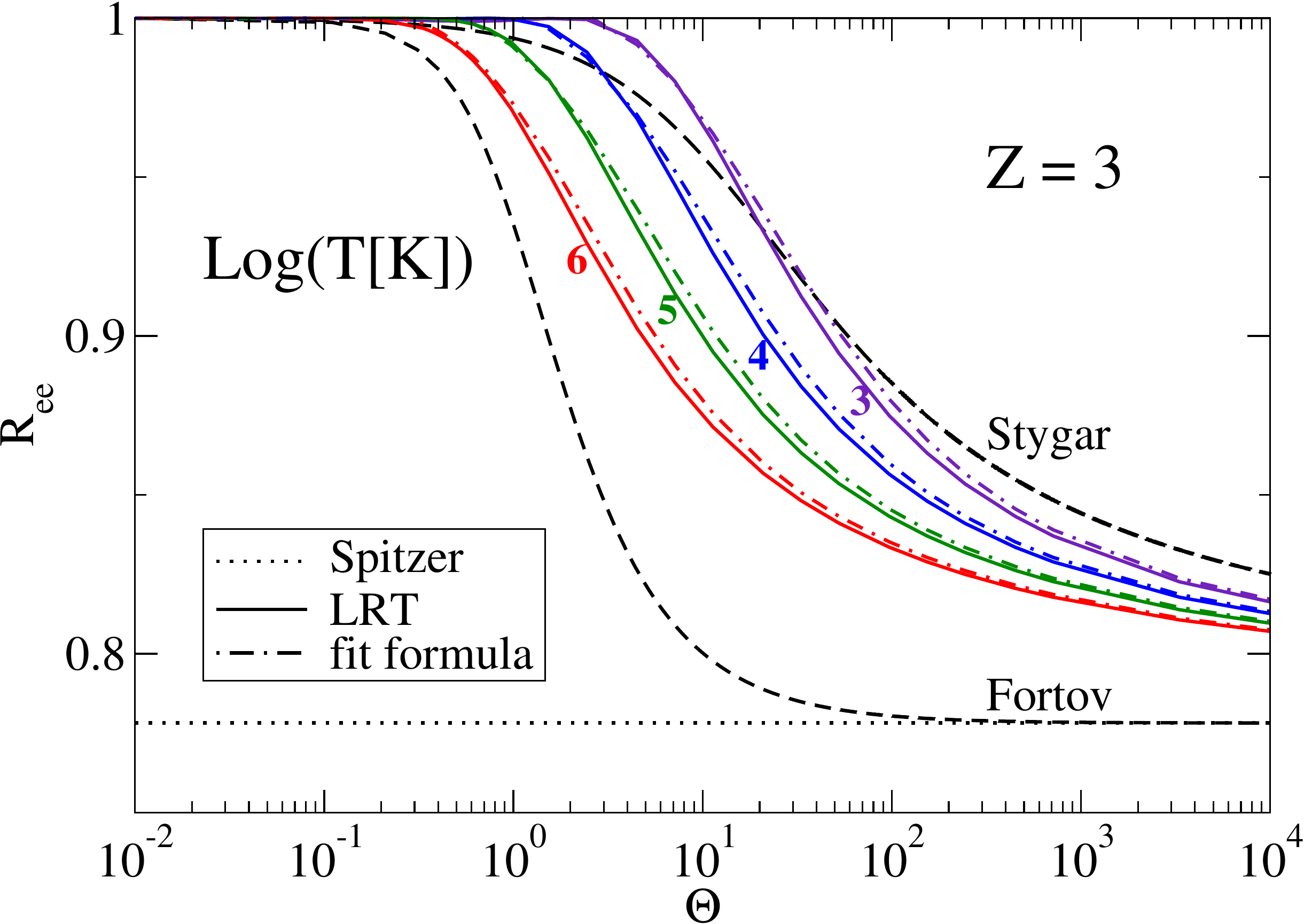}
 \hspace*{5mm}
\end{center}
 \caption{(Color online) Correction factor $R_{\rm ee}$ of the conductivity due to $e-e$ collisions
 as function of the degeneracy parameter $\Theta$ for $Z=2$ (left panel) and $Z=3$ (right panel) for different 
 temperatures $T=(10^3, 10^4, 10^5, 10^6)$~K; same notation as in Fig.~\ref{fig:TZ1}.}
\label{fig:TZ2}
\end{figure}

The fit formula Eq.~(\ref{eq:ReeFit}) for the correction factor $R_{\rm ee}$ is now compared with the  numerical evaluation using the expressions for the correlation functions according to (\ref{deiLBg}, \ref{deeLBg}) in App.~\ref{app0} 
in Figs.~\ref{fig:TZ1} ($Z=1$) and \ref{fig:TZ2} ($Z=2,3$). 

\section{Broadening of the $\delta$-function}
\label{subsdeltaeta}

Another topic is the broadening of the $ \delta$-function to make a smooth transition in the static case  ($\omega \to 0$ ).  
For the application of the Kubo-Greenwood formula given as Eq.~(\ref{KG1}), Desjarlais \textit{et al.}~\cite{Desjarlais02} pointed out: 
"In practice, because of the finite simulation volume
and resulting discrete eigenvalues, the $ \delta$-function must
be broadened. We use a Gaussian broadening of the
$\delta$-function that is as small as feasible without recovering the
local oscillations in the optical conductivity resulting from the discrete band structure".

To discuss expression (\ref{KG1}), we consider a finite value for $\eta$,
\begin{equation}
\delta_\eta(z) = \frac{\eta}{\eta^2+z^2}.
\end{equation}
The finite width of the $\delta$ function can be interpreted as 
an additional damping to overcome the 
level spacing due to the finite volume with periodic boundary conditions. 
The limit $\eta \rightarrow 0$ can be taken only in the final expressions, summing up all orders of perturbation expansion. Expanding with respect to interaction $\hat V$, the van Hove limit ($\hat V^2/\eta \to 0$) has to be taken, see \cite{HC75}. Therefore, for finite $\eta$ a perturbation expansion of (\ref{KG1}) can be performed.

With the perturbation expansion
\begin{equation}
\langle k_1|{\bf \hat p}|k_2 \rangle = \hbar {\bf k}_1 \delta_{k_1,k_2} +\frac{\langle k_1|\hat V|k_2 \rangle}{E_1-E_2} (\hbar {\bf k}_1- \hbar {\bf k}_2)
\end{equation}
we have with ${\bf k}_2={\bf k}_1+{\bf q}$ and $\langle k_1|\hat V|k_2 \rangle=V_q$
\begin{equation} \label{xxx}
{\rm Re}\, \sigma^{\rm KG}(0) = \frac{\pi e^2 \hbar}{3 m^2 \Omega}\sum_{k,q} \frac{\partial f(E_k)}{\partial E_k}
\left({\bf k} \delta_{q,0}+\frac{V_q}{E_k-E_{k+q}}{\bf q}+\dots\right)^2  \frac{\eta}{\eta^2+(E_k-E_{k+q})^2}.
\end{equation}
Considering the screened interaction with uncorrelated ions in the nondegenerate case, 
$V_q^2 = N_{\rm ion} Z^2 e^4/[\epsilon_0 \Omega (q^2 +\kappa^2)]^2$, Eq. (\ref{xxx}) leads to
\begin{equation}
{\rm Re}\, \sigma^{\rm KG}(0) = \frac{\pi e^2 \hbar \beta}{3 m^2 }\int \frac{d^3 k}{(2 \pi)^3} f(E_k)
\left(k^2 \frac{1}{\eta}+\int\frac{d^3 q}{(2 \pi)^3}\frac{n_{\rm ion} Z^2 e^4}{[\epsilon_0 (q^2 +\kappa^2)]^2 (E_k-E_{k+q})^2}
q^2 \frac{\eta}{\eta^2+(E_k-E_{k+q})^2}+\dots\right).
\end{equation}
Before the last term is reinterpreted as a $\delta$-function, we estimate the denominator $E_k-E_{k+q}$ 
by the broadening parameter $\eta$ of the $\delta_\eta$ function so that
\begin{equation}
{\rm Re}\, \sigma^{\rm KG}(0) = \frac{\pi e^2 \hbar \beta}{3 m^2 }\int \frac{d^3 k}{(2 \pi)^3} f(E_k)
k^2 \tau^{\rm KG}(k)+\dots
\end{equation}
with
\begin{eqnarray}
\tau^{\rm KG}(k) &=&  \frac{1}{\eta}+ \frac{1}{k^2}\int\frac{d^3 q}{(2 \pi)^3}\frac{n_{\rm ion} Z^2 e^4}{[\epsilon_0 (q^2 +\kappa^2)]^2 (E_k-E_{k+q})^2}
q^2 \frac{\eta}{\eta^2+(E_k-E_{k+q})^2}+\dots
\nonumber \\ &&
= \frac{1}{\eta}+ \frac{1}{k^2}\frac{n_{\rm ion} Z^2 e^4 }{\epsilon^2_0}
\int_0^\infty\frac{d q}{(2 \pi)^2}\frac{q^4}{(q^2 +\kappa^2)^2 }
\left(\frac{m}{\hbar^2kq}\right)^3 \int_{-1}^1 dz \frac{\eta m/kq}{(\eta m/kq)^2+(z+q/2k)^2}\frac{1}{(z+q/2k)^2}
\nonumber \\ &&
=\frac{1}{\eta}+ \frac{1}{k^2}\frac{n_{\rm ion} Z^2 e^4 }{\epsilon^2_0}
\int_0^\infty\frac{d q}{(2 \pi)^2}\frac{q^4}{(q^2 +\kappa^2)^2 }\left(\frac{m}{\hbar^2kq}\right)^3 
\left[ \frac{\hbar^2kq}{\eta m}\frac{2}{1-(q/2k)^2}
+\pi \left(\frac{\hbar^2kq}{\eta m}\right)^2 \right]
\nonumber \\ &&
=\frac{1}{\eta}+ \frac{1}{\eta^2}\frac{1}{k^3}\frac{n_{\rm ion} Z^2 e^4 m \pi}{\epsilon^2_0 \hbar^2 }\int_0^{2 k}\frac{d q}{(2 \pi)^2}
\frac{q^3}{(q^2 +\kappa^2)^2 } +{\cal O}\left(\frac{e^4}{\eta}\right)\,.
\end{eqnarray}
In principle, one has to sum the leading divergent terms $\propto (1/\eta) \left(e^4/\eta\right)^n$. We give here only the first contributions,
\begin{equation}
\frac{1}{\eta}+ \frac{1}{\eta^2}A+\dots = \frac{1}{\eta}\left[1+ \frac{1}{\eta}A+\dots\right] = \frac{1}{\eta}\frac{1}{1- \frac{1}{\eta}A+\dots}.
\end{equation}
Now the limit $\eta \to 0$ can be performed with the result $-1/A$.

For comparison, see \cite{Reinholz12}, with the golden rule for the transition rates and 
$S(q)\approx 1 \rightarrow |V_{\textrm{ei}}(q)|^2\approx V_{q}^{2} $, 
the energy dependent relaxation time can be calculated
 \begin{eqnarray} \label{reltime1}
  \frac{1}{\tau_k} &=&
 -\frac{2 \pi}{\hbar}\sum_{q} V^2_{q} \delta(E_k-E_{k+q})\frac{\bf E \cdot \bf q }{\bf E \cdot \bf k }.
 \end{eqnarray}
 The $\bf q$ integral in \rf{reltime1} can be performed using spherical coordinates where $\bf k$ is in $z$ direction, $\bf E$ in the $x-z$ plane.
It is convergent only in the case of a  screened Coulomb potential. Using the 
statically screened Debye potential 
$V_{q}=e^2/\{\epsilon_0 \Omega_0 (q^2+\kappa^2_{\rm D})\} \, , \, \kappa^2_{\rm D}=\beta n_{\rm e} e^2 /\epsilon_0 $, 
we find the energy dependent collision frequency
\begin{equation} \label{coulomblog}
\nu_k=\tau_k^{-1}=n_{\rm e} \frac{Z e^4}{4 \pi \epsilon_0^2} \frac{m}{\hbar^3 k^3}\left( \ln \sqrt{1+b}-\frac{1}{2} \frac{b}{1+b}\right),
\end{equation}
with $b= 4 k^2 /\kappa^2_{\rm D}$ in the Coulomb logarithm. 
The static conductivity is determined as
\begin{eqnarray} \label{jel}
\sigma_{\textrm{dc}}^{\rm Lorentz} &=& 
\frac{e^2 \hbar^2}{m^2}\beta \frac{1}{ \Omega_0} \sum_k \, k_E^2 \, \tau_k \,f_k(1-f_k) 
= \epsilon_0 \omega^2_{\textrm{pl}} \tau^{\rm Lorentz}=\frac{e^2 n_{\rm e}}{m\, \nu^{\rm Lorentz}}\,.
\end{eqnarray}
We introduce the average relaxation time $\tau^{\rm Lorentz}$ 
and the static collision frequency $\nu^{\rm Lorentz}=1/\tau^{\rm Lorentz}$.
The approach can also be applied for a pseudo-potential describing the $e-i$ interaction 
and an ion structure factor describing the ion configuration.
The Lorentz model is solved if using the relaxation time ansatz. It corresponds to the Brooks-Herring result 
where the semiconductor conductivity for the screened electron-hole interaction is considered. 


\end{document}